\documentclass[longauth]{aa}

\usepackage{graphicx}
\usepackage{natbib}
\usepackage{flexisym}
\usepackage{letltxmacro}
\usepackage{amsmath}
\usepackage{xcolor}
\LetLtxMacro\FlexiSymTextPrime\textprime
\DeclareTextCommandDefault{\textprime}{\FlexiSymTextPrime}   
\usepackage{subcaption}  
\usepackage{float}      
\usepackage[utf8]{inputenc}
\usepackage{graphicx}
\usepackage{caption}
\usepackage{subcaption}

\usepackage{hyperref}
\hypersetup{colorlinks=True, linkcolor=blue, citecolor=blue}

\usepackage{txfonts,textcomp}
\begin{document}

   \title{PAHSPECS: Polycyclic aromatic hydrocarbon properties at cosmic noon with JWST/MIRI MRS}

    \author{C. M. Lofaro\inst{1,2}, T. Díaz Santos\inst{2,3}, I. Shivaei\inst{4}, L. A. Boogaard\inst{5}, K. Sandstrom\inst{6}, F. R. Donnan\inst{6}, R. Fernández Aranda\inst{4}, P. G. P\'erez-Gonz\'alez\inst{4}, M. Aravena\inst{7,8}, R. Decarli\inst{9}, F. Galliano\inst{10}, H. Inami\inst{11}, G. \"Ostlin\inst{12}, G. Popping\inst{13},  P. P. van der Werf\inst{5}, F. Walter\inst{14,15}}
    
    \institute{Department of Physics, University of Crete, 70013, Heraklion, Greece\\
              \email{clofaro@ia.forth.gr}
    \and
    Institute of Astrophysics, Foundation for Research and Technology–Hellas (FORTH), Heraklion, GR-70013, Greece
    \and
    School of Sciences, European University Cyprus, Diogenes street, Engomi, 1516 Nicosia, Cyprus
    \and
    Centro de Astrobiología (CAB), CSIC-INTA, Carretera de Ajalvir km 4, Torrejón de Ardoz, E-28850, Madrid, Spain
    \and
    Leiden Observatory, Leiden University, PO Box 9513, 2300 RA Leiden, The Netherlands
    \and
    Department of Astronomy \& Astrophysics, University of California San Diego, 9500 Gilman Drive, San Diego, CA 92093, USA
    \and
    Instituto de Estudios Astrof\'{\i}sicos, Facultad de Ingenier\'{\i}a y Ciencias, Universidad Diego Portales, Av. Ej\'ercito 441, Santiago, Chile
    \and
    Millenium Nucleus for Galaxies (MINGAL)
    \and
    INAF–Osservatorio di Astrofisica e Scienza dello Spazio, via Gobetti 93/3, I-40129, Bologna, Italy,
    \and
    Université Paris-Saclay, Université Paris Cité, CEA, CNRS, AIM, 91191, Gif-sur-Yvette, France
    \and
    Hiroshima Astrophysical Science Center, Hiroshima University, 1-3-1 Kagamiyama, Higashi-Hiroshima, Hiroshima 739-8526, Japan
    \and
    Department of Astronomy, Oscar Klein Centre, AlbaNova University Center, Stockholm University, 10691 Stockholm, Sweden
    \and
    European Southern Observatory, Karl-Schwarzschild-Strasse 2, D-85748, Germany
    \and
    Max Planck Institut f\"ur Astronomie, K\"onigstuhl 17, D-69117 Heidelberg, Germany
    \and
    California Institute of Technology, Pasadena, CA 91125, USA}

  \date{Received date /
    Accepted date }
   
   \titlerunning{PAHSPECS}
   \authorrunning{Lofaro C. M. et al.}

\abstract
   {The epoch of "cosmic noon" ($z{\sim}1$--3) marks the peak of the cosmic star-formation rate density and represents a phase in which dust-obscured star formation dominates galaxy growth. Mid-infrared (MIR) spectroscopy provides a powerful probe of the interstellar medium (ISM) during this period through emission from polycyclic aromatic hydrocarbons (PAHs), whose vibrational band ratios trace the charge state, the size distribution of the PAH population and the local radiation field conditions.}
   {In this work, we characterize the PAH properties of star-forming galaxies at $z{\sim}1.1$ and investigate how their PAH luminosities and band ratios relate to global galaxy properties such as infrared luminosity, star-formation rate (SFR), and specific star-formation rate (sSFR). By comparing these systems to local luminous infrared galaxies (LIRGs), we seek to assess whether the nature of PAH emission at cosmic noon differ systematically from those in the nearby Universe.}
   {We analyze observations from the PAHSPECS survey, consisting of JWST/MIRI Medium Resolution Spectroscopy (MRS) observations of five star-forming galaxies drawn from the ALMA Spectroscopic Survey (ASPECS) in the Hubble Ultra Deep Field (HUDF). Integrated spectra are extracted using wavelength-dependent apertures and modeled with the \texttt{CAFE} spectral fitting code where we also incorporated ancillary photometry to better constrain the dust emission. Stellar mass and SFR were derived via spectral energy distribution fitting with \textit{Prospector}.}
   {Compared to local LIRGs, most of the PAHSPECS sources exhibit higher 6.2/7.7 and lower 11.3/7.7 ratios. In the framework of PAH models, these offsets suggest that the ionized PAH component is weighted toward smaller grains relative to the nearby systems. The 3.3/11.3 ratio is less well constrained, since the 3.3~$\mu$m feature is detected in only two sources, likely due to enhanced processing of the smallest PAH carriers in harder radiation fields. Although no firm conclusion can be drawn about the size distribution of the neutral PAHs, our measurements remain compatible with a population weighted toward larger neutral PAHs. Within the PAHSPECS sample, 11.3/7.7 increases with both sSFR and star-formation surface density, while the 6.2/7.7 ratio decreases with increasing sSFR, consistent with the preferential processing or destruction of small and ionized PAH carriers. The AGN-hosting source ASPECS-15 stands out with the lowest 6.2/7.7 ratio and highest sSFR in the sample, suggesting a reduced contribution from small PAHs, potentially due to AGN activity. The 7.7~$\mu$m luminosity follows the local $L_{7.7}$--SFR relation, supporting its use as a star-formation tracer at $z{\sim}1$.}
  {These results suggest that PAH emission at cosmic noon is shaped by different ISM conditions than in nearby starburst galaxies, likely reflecting the more intense radiation-field conditions at the cosmic SFR peak. While the 7.7~$\mu$m feature remains a robust tracer of star formation, the PAH band ratios indicate systematic differences in the dust properties of $z{\sim}1$ main-sequence galaxies relative to local LIRGs.}


   \maketitle
%

\section{Introduction}
The epoch commonly referred to as cosmic noon, spanning approximately $z{\sim}1$--3, corresponds to the peak of the cosmic star-formation rate density and to a phase in which dust-obscured star formation dominates the global star-formation budget \citep[e.g.,][]{Madau14,Zavala21}. Galaxies at these redshifts are typically gas-rich and actively forming stars, with molecular-gas fractions and star-formation efficiencies that differ systematically from those observed in the local Universe, as established by CO- and dust-based surveys and scaling relations \citep{Tacconi13,Scoville17,Tacconi18, Tacconi20, Aravena19, Aravena20, Boogaard20}. As a result, understanding how stars are formed during this epoch requires a detailed characterization of the physical conditions of the interstellar medium (ISM) in which they are embedded.

The ISM of star-forming galaxies can be probed effectively at mid-infrared (MIR) wavelengths, where emission from polycyclic aromatic hydrocarbons (PAHs) traces the interface between molecular gas and the regions exposed to ultraviolet (UV) radiation. PAHs are large aromatic molecules, representing only a small fraction of the total dust mass, that are stochastically heated by UV photons and re-emit their energy through a series of prominent MIR features, with the strongest bands commonly observed at 3.3, 6.2, 7.7, 8.6, 11.3, and 17~$\mu$m \citep{Leger84, Allamandola85, Allamandola89, Tielens08, Galliano08}. The relative strengths of the PAH bands depend on both the PAH charge state, size distribution, and the radiation field spectrum. In particular, the 3.3~$\mu$m and 11.3~$\mu$m features are generally stronger in neutral PAHs, while the 6.2~$\mu$m and 7.7~$\mu$m bands become relatively more prominent in ionized PAHs \citep{Allamandola99, Galliano08, Boersma16, Boersma18, Maragkoudakis18, Draine21, Rigopoulou21,  Maragkoudakis22}. In addition, PAH emission at longer wavelengths is preferentially associated with larger PAH molecules, whereas shorter-wavelength features tend to arise from smaller PAHs, reflecting a systematic increase of characteristic PAH size with wavelength \citep{Draine07,Tielens08,Galliano08,Draine21}. Since the ionization balance depends on the intensity and hardness of the radiation field (as well as on the local electron density and gas conditions), and since the radiation field spectrum also affects the temperature distribution of PAHs, PAH band ratios provide a physically motivated diagnostic of the UV radiation environment, particularly in photo-dissociation regions (PDRs), as they can vary with the radiation field spectrum even at fixed size and charge \citep{Draine07,Draine21}.

In nearby star-forming galaxies, PAH emission is found to correlate with infrared luminosity (L$_{IR}$) and star-formation rate (SFR), and has therefore been widely used as an indirect tracer of star formation \citep{Peeters04, Smith07, Pope08, Shipley16}. For instance, the Great Observatories All-sky LIRG Survey (GOALS) is a reference sample of luminous infrared galaxies (LIRGs) with uniform MIR spectroscopy \citep{Armus09} that has enabled detailed analyses of PAH equivalent widths, band ratios, and their connection to starburst and active galactic nuclei (AGN) activity \citep{Sabrina13, Sabrina14}. In addition, combined \textit{AKARI}+\textit{Spitzer} spectroscopy for GOALS galaxies has been used to further quantify the contribution of AGN-heated dust to the MIR continuum and its impact on PAH observables \citep{Inami18}. PAH properties are now widely employed to discriminate between star-formation and AGN-dominated systems, since AGN can reduce the apparent PAH strength through a strong hot-dust continuum (dilution) and, in some environments, through modification or destruction of PAH carriers (\citealp{Leach86, Leger89, Voit92, Sabrina13, Sabrina14, Inami18, McKinney21, Garcia-Bernete22};
\textcolor{blue}{Lofaro et al.\ submitted 2026a}). This phenomenon is well documented in the literature, particularly for strong AGN with minimal obscuration, which produce nearly featureless spectra, predominantly attributed to hot dust emission \citep{Stern05, Alonso_Herrero06, Donley07, Li07, Tanio10, Mullaney11}.

MIR spectroscopy at high redshift has established that strong PAH bands (e.g., 6.2, 7.7, and 11.3~$\mu$m) are commonly present in $z{\sim}1{-}2$ star-forming systems \citep[e.g.,][]{Murata14a,   Murata14b, Liu25} and that their luminosities correlate with L$_{\mathrm{IR}}$ in ways broadly consistent with the local PAH--L$_{\mathrm{IR}}$ relation \citep{Pope08, Menendez09}. Building on locally calibrated relations, \citet{Shipley16} derived empirical PAH--SFR conversions and demonstrated that PAH luminosities can serve as quantitative tracers of star formation in high-redshift galaxies, while also emphasizing intrinsic scatter and metallicity dependence. Using MOSDEF galaxies \citep{Kriek15}  at 1.37$<$z$<$2.61, \citet{Shivaei17} showed that the ratio L$_{7.7}/L_{\mathrm{IR}}$ depends strongly on gas-phase metallicity and ionization state, with suppressed PAH emission observed in lower-metallicity systems. Similarly, in the local Universe, metallicity has been identified as a primary driver of the PAH equivalent width \citep{Galliano18}. Intermediate-redshift studies have compared PAH-strength indicators at $z{\sim}0.8$ with both local and $z{\sim}2$ samples and demonstrated that the PAH-to-L$_{\mathrm{IR}}$ ratio varies across the star-forming population rather than following a single universal conversion \citep{Murata15}.

The advent of JWST has dramatically advanced studies of galaxies at cosmic noon by providing a combination of angular resolution, sensitivity, and integral-field spectroscopy that was not accessible with previous mid-infrared facilities. In particular, JWST enables the simultaneous detection of both the 3.3 and 11.3~$\mu$m PAH features in galaxies out to $z{\sim}1.2$, representing a major step forward in the characterization of PAH properties at cosmic noon by directly probing both neutral and larger PAH populations within the same systems.
Using JWST photometry, \citet{Shivaei24} find a strong correlation between the PAH dust fraction ($q_{\mathrm{PAH}}$) and stellar mass in galaxies at $z{\sim}0.7$--2, and show that the relation between $q_{\mathrm{PAH}}$ and gas-phase metallicity closely follows that observed in the local Universe, with $q_{\mathrm{PAH}}$ remaining approximately constant above $Z\gtrsim0.5\,Z_\odot$ and dropping sharply at lower metallicities, implying that PAH emission can reliably trace obscured star formation at cosmic noon in metal-enriched systems. JWST/MIRI Low Resolution Spectroscopy (LRS) further enables direct comparisons between high-redshift and local systems: based on a sample of 37 galaxies at $z{=}0.65$--2.46, \citet{McKinney25} report systematically enhanced 11.3/3.3~$\mu$m PAH ratios relative to nearby galaxies, suggesting differences in the characteristic PAH populations at cosmic noon. Resolved JWST studies of nearby galaxies illustrate the diagnostic power of PAHs when spatial information is available. In particular, \citet{Lai22,Lai23} show that PAH emission, including the 3.3~$\mu$m feature, is strongly reduced in regions dominated by AGN activity while remaining prominent in surrounding star-forming regions, highlighting how AGN can influence PAHs emission.
At the same time, PAH luminosities show a tight empirical connection with molecular gas tracers over a broad redshift range, indicating that PAHs remain closely linked to the cold gas reservoir that fuels star formation even when their efficiency as star-formation tracers varies \citep{Cortzen19, Leroy23, ShivaeiBoo24}. These results suggest that PAH emission encodes critical information about ISM conditions and star-formation mode, further motivating detailed comparisons between cosmic-noon galaxies and nearby star-forming systems.

In this work, we analyze JWST/MIRI Medium Resolution Spectroscopy (MRS) observations of a complete sample of five star-forming galaxies at $z{\sim}1.1$ obtained as part of the PAHSPECS program (GO~5279; co-PIs: Shivaei, Diaz-Santos, Boogaard), specifically focusing on PAHs integrated measurements. Companion papers will present the selection, general properties and basic measurements of the sample \citep{Shivaei26}; perform a spatially resolved analysis of the features \citep{Donnan26}; and relate the PAHs to the gas and dust properties of the galaxies \citep{Boogaard26}

This paper is organized as follows. Sect.~\ref{sample and JWST reduction} presents the observations and data reduction of the galaxy sample. Sect.~\ref{spectral extraction} details the spectral extraction and emission-line fitting procedure. In Sect.~\ref{results} we present the integrated measurements of the PAHs and PAH ratios, and relate them to integrated properties of the galaxies such as LIR, SFR and sSFR. The interpretation and implications for comic noon galaxies are  discussed in Sect.~\ref{discussion}. Finally, we summarize our conclusions in Sect.~\ref{conclusions}.

Throughout this paper, we assume a flat $\Lambda$CDM cosmology with $H_0 = 67.4$ km s$^{-1}$ Mpc$^{-1}$, $\Omega_{\rm m} = 0.315$, and $\Omega_{\Lambda} = 0.685$ \citep{Plank20}.

\begin{table*}[]
\centering
\footnotesize
\caption{PAHSPECS Sample Properties}
\label{tab:pahspecs_sample_props}
\resizebox{\textwidth}{!}{%
\begin{tabular}{lccccccccc}
\hline\hline
Name$^{(1)}$ & 3mm ID / 1mm ID$^{(2)}$ & RA$^{(3)}$ & Dec$^{(4)}$ & $\rm z_{spec}$$^{(5)}$ & $r_{e}^{15\,\mu\mathrm{m}}$$^{(6)}$ & $\tau_{9.7}$$^{(7)}$ & $\log L_{\rm IR}$$^{(8)}$ & SFR$^{(9)}$ & $\log M_{*}$$^{(10)}$ \\
& & (J2000) & (J2000) & & ($\prime\prime$) & & ($\rm L_{\odot}$) & ($\rm M_{\odot}\,yr^{-1}$) & ($\rm M_{\odot}$) \\
\hline
  ASPECS-6 & 3mm.06 / 1mm.C16 & 03:32:39.88 & $-27$:47:15.24 & 1.0952 & 0.63 & $0.5 \pm 0.3$ & $11.43^{+0.03}_{-0.02}$ & $28.9^{+1.9}_{-2.4}$ & $10.82^{+0.05}_{-0.05}$ \\
  ASPECS-11 & 3mm.11 / -- & 03:32:39.82 & $-27$:46:53.79 & 1.0964 & 0.41 & $0.9 \pm 0.3$ & $10.80^{+0.09}_{-0.06}$ & $6.8^{+0.9}_{-0.7}$ & $10.44^{+0.04}_{-0.04}$ \\
  ASPECS-14 & 3mm.14 / 1mm.C25 & 03:32:34.86 & $-27$:46:40.77 & 1.0982 & 0.40 & $0.3 \pm 0.2$ & $11.44^{+0.03}_{-0.03}$ & $22.4^{+2.4}_{-1.9}$ & $10.65^{+0.06}_{-0.06}$ \\
  ASPECS-15 & 3mm.15 / 1mm.C12 & 03:32:36.48 & $-27$:46:31.95 & 1.0931 & 0.63 & $1.9 \pm 0.7$ & $11.58^{+0.06}_{-0.06}$ & $40.5^{+6.9}_{-6.9}$ & $10.26^{+0.06}_{-0.06}$ \\
  ASPECS-C20 & -- / 1mm.C20 & 03:32:35.78 & $-27$:46:27.83 & 1.0963 & 0.47 & $0.8 \pm 0.4$ & $11.18^{+0.05}_{-0.04}$ & $10.5^{+1.7}_{-2.4}$ & $10.98^{+0.06}_{-0.05}$ \\
\hline
\end{tabular}
}
\begin{minipage}{0.95\textwidth}
\vspace{0.5cm}
\footnotesize
\textbf{Notes.} Columns: (1) Source name. (2) 3~mm CO ID and 1~mm continuum ID. (3)--(4) Coordinates from our simultaneous 15 $\mu$m image. (5) CO redshift. (6) Effective radius at 15 $\mu$m. (7) Silicate optical depth at 9.7~$\mu$m. (8), (9), (10) IR luminosity, SFR and stellar mass from SED fitting.
\end{minipage}
\end{table*}

\begin{figure*}[t]
	\centering
		\includegraphics[width=1\textwidth]{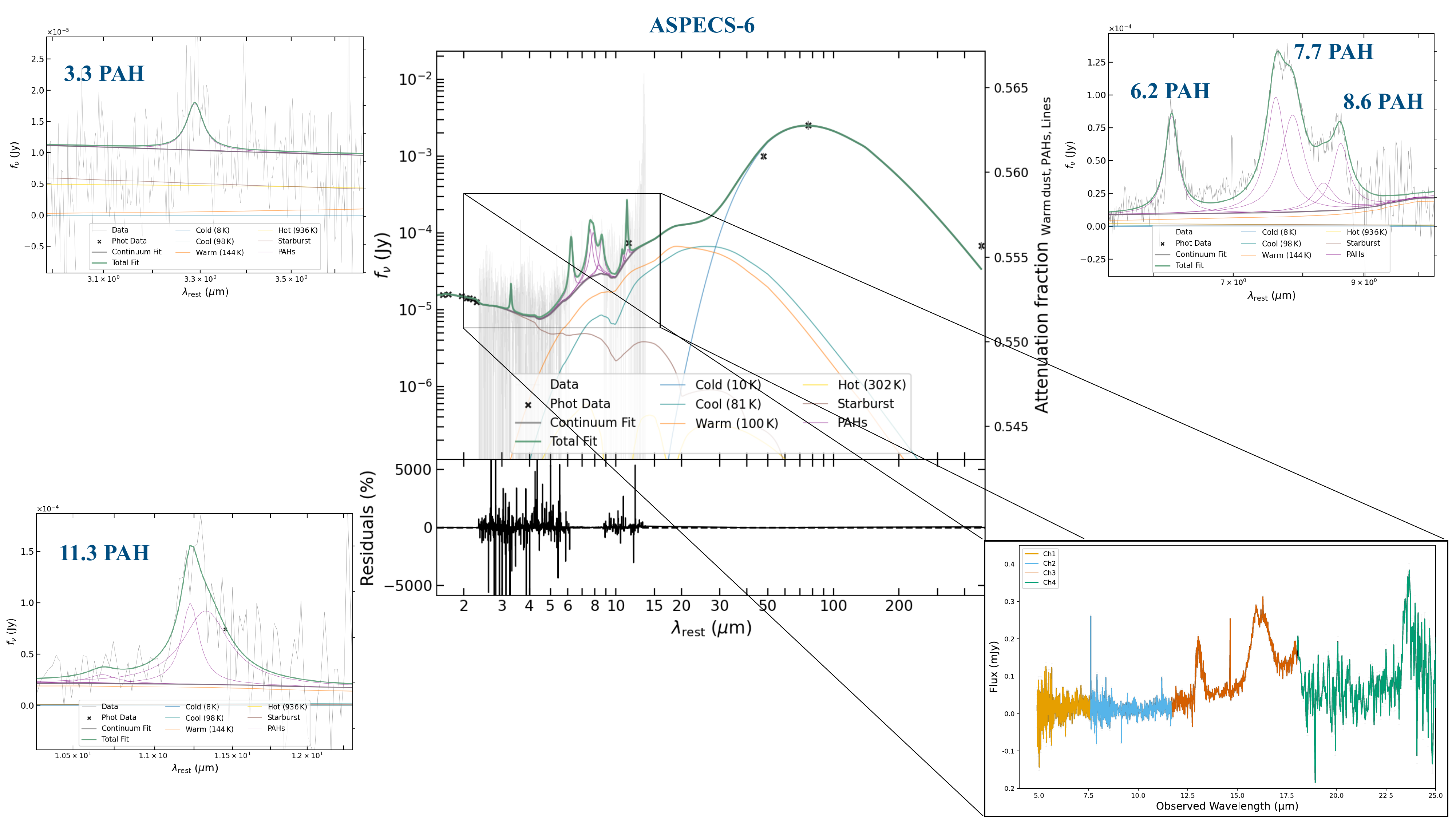}
		\caption{ASPECS-6: Spectral fit using the full JWST/MIRI MRS wavelength coverage (light grey), together with the available photometric data (black crosses, see Sect.~\ref{spectral extraction}). The green line represents the total fit. The dark gray line shows the underlying continuum made of the cold dust (blue), cool dust (cyan), warm dust
        (orange), and hot dust component (not shown). The purple lines are the PAHs.}
        \label{fig: CAFE fit aspecs6}
\end{figure*}

\begin{table*}
\centering
\caption{PAH luminosities of the PAHSPECS galaxies. Left: observed (uncorrected) values. Right: extinction-corrected values. $1\sigma$ upper limits are reported for the 3.3~$\mu$m feature in ASPECS-11, ASPECS-15, and ASPECS-C20.}
\label{tab:pah_fluxes}
\renewcommand{\arraystretch}{1.2}
\resizebox{\textwidth}{!}{%
\begin{tabular}{lccccc|ccccc}
\hline \hline
 & \multicolumn{5}{c|}{Observed} & \multicolumn{5}{c}{Extinction-corrected} \\
 & \multicolumn{5}{c|}{$\log L_{\rm PAH}/L_{\odot}$} & \multicolumn{5}{c}{$\log L_{\rm PAH}/L_{\odot}$} \\
Name & 3.3 & 6.2 & 7.7 & 8.6 & 11.3 & 3.3 & 6.2 & 7.7 & 8.6 & 11.3 \\
\hline
ASPECS-6 & $8.73 (\pm 8.40)$ & $9.60 (\pm 8.96)$ & $10.05 (\pm 9.21)$ & $9.40 (\pm 8.88)$ & $9.24 (\pm 8.93)$ & $8.86 (\pm 8.53)$ & $9.66 (\pm 9.02)$ & $10.11 (\pm 9.26)$ & $9.49 (\pm 8.96)$ & $9.37 (\pm 9.07)$ \\
ASPECS-11 & ${<}8.12$ & $8.81 (\pm 8.13)$ & $9.31 (\pm 8.09)$ & $8.63 (\pm 8.29)$ & $8.46 (\pm 8.00)$ & ${<}8.12$ & $8.87 (\pm 8.19)$ & $9.37 (\pm 7.14)$ & $8.71 (\pm 8.37)$ & $8.59 (\pm 8.27)$ \\
ASPECS-14 & $8.00 (\pm 7.63)$ & $9.37 (\pm 8.50)$ & $9.86 (\pm 8.85)$ & $9.27 (\pm 8.44)$ & $9.26 (\pm 8.64)$ & $8.04 (\pm 7.66)$ & $9.39 (\pm 8.52)$ & $9.88 (\pm 8.87)$ & $9.29 (\pm 8.47)$ & $9.30 (\pm 8.68)$ \\
ASPECS-15 & ${<}8.06$ & $8.70 (\pm 7.92)$ & $9.61 (\pm 7.84)$ & $8.40 (\pm 7.46)$ & $8.69 (\pm 8.11)$ & ${<}8.06$ & $8.84 (\pm 8.05)$ & $9.72 (\pm 7.96)$ & $8.55 (\pm 7.62)$ & $8.94 (\pm 8.36)$ \\
ASPECS-C20 & ${<}8.22$ & $9.22 (\pm 8.44)$ & $9.70 (\pm 8.05)$ & $9.04 (\pm 8.67)$ & $8.60 (\pm 7.86)$ & ${<}8.22$ & $9.28 (\pm 8.49)$ & $9.75 (\pm 8.10)$ & $9.12 (\pm 8.75)$ & $8.72 (\pm 7.99)$ \\
\hline
\end{tabular}}
\end{table*}

\section{Observations and Data Reduction}
\label{sample and JWST reduction}

The five targets observed in the PAHSPECS JWST program were drawn from the flux-limited 1.2~mm and 3~mm observations of the ALMA Spectroscopic Survey \citep[ASPECS;][]{Walter16, Decarli19, Aravena20}, tracing CO(2--1) and dust continuum in galaxies spanning a wide range of redshifts, from $z{\approx}$0.5--4 \citep{Boogaard19,Boogaard20,Decarli19, Gonzalez19, Gonzalez20} in the Hubble Ultra Deep Field (HUDF).
The PAHSPECS sources are main-sequence star-forming galaxies at $z{\sim}1$ and, as they are drawn from the CO-flux-complete ASPECS sample, probe a representative population of star-forming galaxies at this epoch \citep{Boogaard19}. 
Table~\ref{tab:pahspecs_sample_props} summarizes the general properties of the sample. A comprehensive overview of the sample selection and derived physical parameters, including those inferred from broadband SED modeling, will be presented in \citep{Shivaei26}.

We obtained observations using the JWST/MIRI MRS, which provides integral field spectroscopy covering the full 5-28 $\mu m$ wavelength range (${\sim}$~2.4--13.3~$\mu m$ rest-frame). The observations were conducted using the SLOWR1 readout mode for enhanced sensitivity. A 4-point dither pattern was employed for source observations, while the dedicated background observations used a 2-point dither pattern to optimize observing efficiency. Simultaneous MIRI imaging was obtained to improve astrometric calibration of the spectroscopic observations. For further details, we refer the reader to \citet{Shivaei26}.

The JWST/MIRI MRS observations were reduced using the official JWST Science Calibration Pipeline (version 1.17.0, \cite{JWST_pipeline}), with the Calibration Reference Data System (CRDS) context \texttt{jwst_1321.pmap}. The reduction followed the standard three-stage pipeline structure: at each stage, we retained the default pipeline configuration while fine-tuning selected steps relevant to cosmic ray rejection, fringing correction, and background subtraction.

In Stage 1, the \texttt{Detector1Pipeline} was applied to the uncalibrated exposures (\texttt{*_uncal.fits}) to perform detector-level corrections, resulting in \texttt{*_rate.fits} files. We activated the cosmic ray shower detection and disabled the flicker noise filtering to prevent unwanted suppression of faint spatial structures. All other steps followed the default pipeline implementation. 

In Stage 2, the \texttt{Spec2Pipeline} was executed on the \texttt{*_rate.fits} files to apply WCS calibration, flat-fielding, fringing correction, and photometric calibration, resulting in the \texttt{*_cal.fits} and \texttt{*_s2d}.fits products. To further mitigate instrumental effects, we enabled residual fringing removal. We also applied the additional targeted cosmic-ray mitigation step to remove residual large-scale artifacts associated with cosmic ray showers, which are not fully corrected by standard jump detection. This procedure significantly reduces correlated noise and improves the extraction of weak spectral features.
We tested different background subtraction strategies using dedicated ASPECS-6 off-source observations as background for all sources. In particular, we compared pixel-level background subtraction, in which the background is removed independently in each spaxel, with the use of a master background that subtracts a field-averaged background. Based on these tests, we adopted the pixel-level approach to avoid spatial smoothing and potential over-subtraction. 

\begin{figure*}[]
    \centering
    \begin{subfigure}[b]{0.24\textwidth}
        \includegraphics[width=\textwidth]{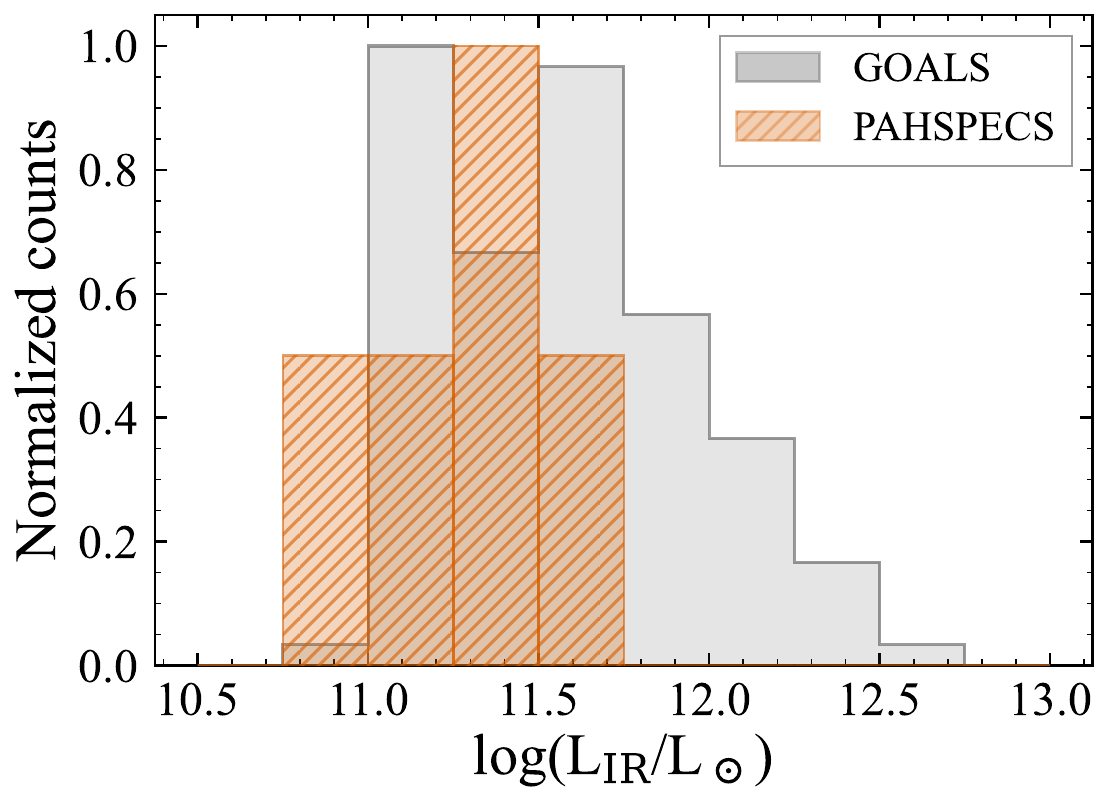}
    \end{subfigure}
    \hfill
    \begin{subfigure}[b]{0.24\textwidth}
        \includegraphics[width=\textwidth]{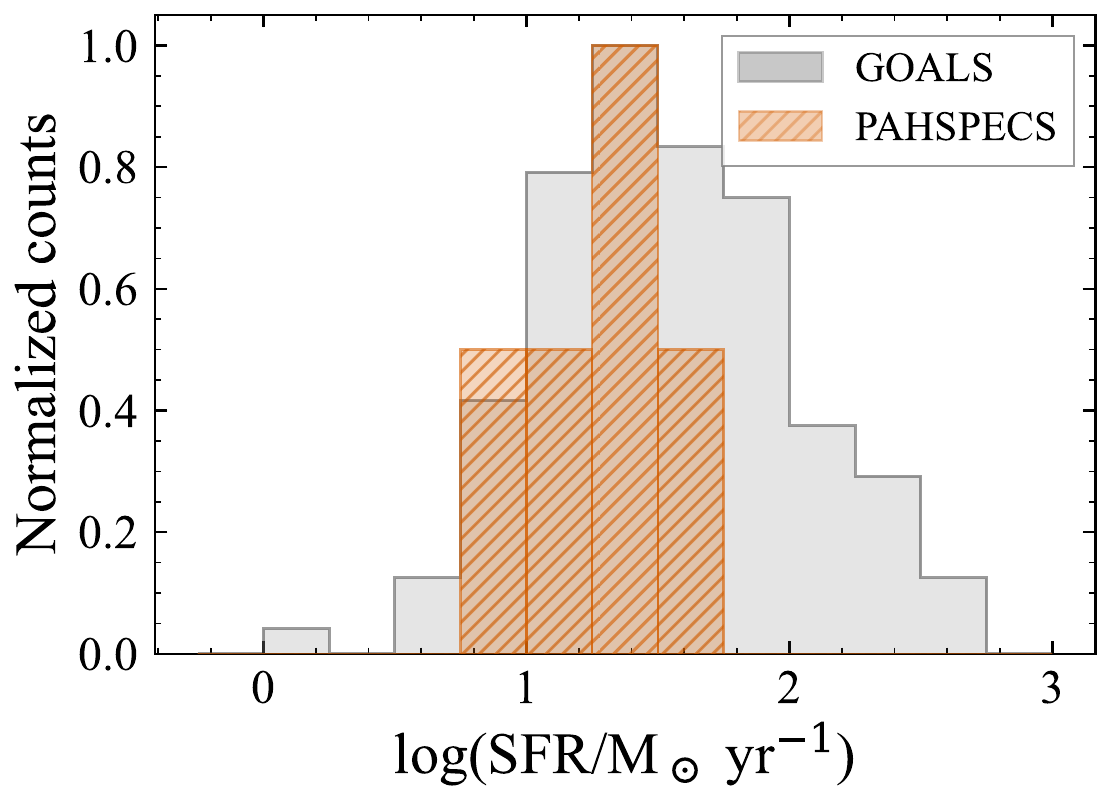}
    \end{subfigure}
    \hfill
    \begin{subfigure}[b]{0.24\textwidth}
        \includegraphics[width=\textwidth]{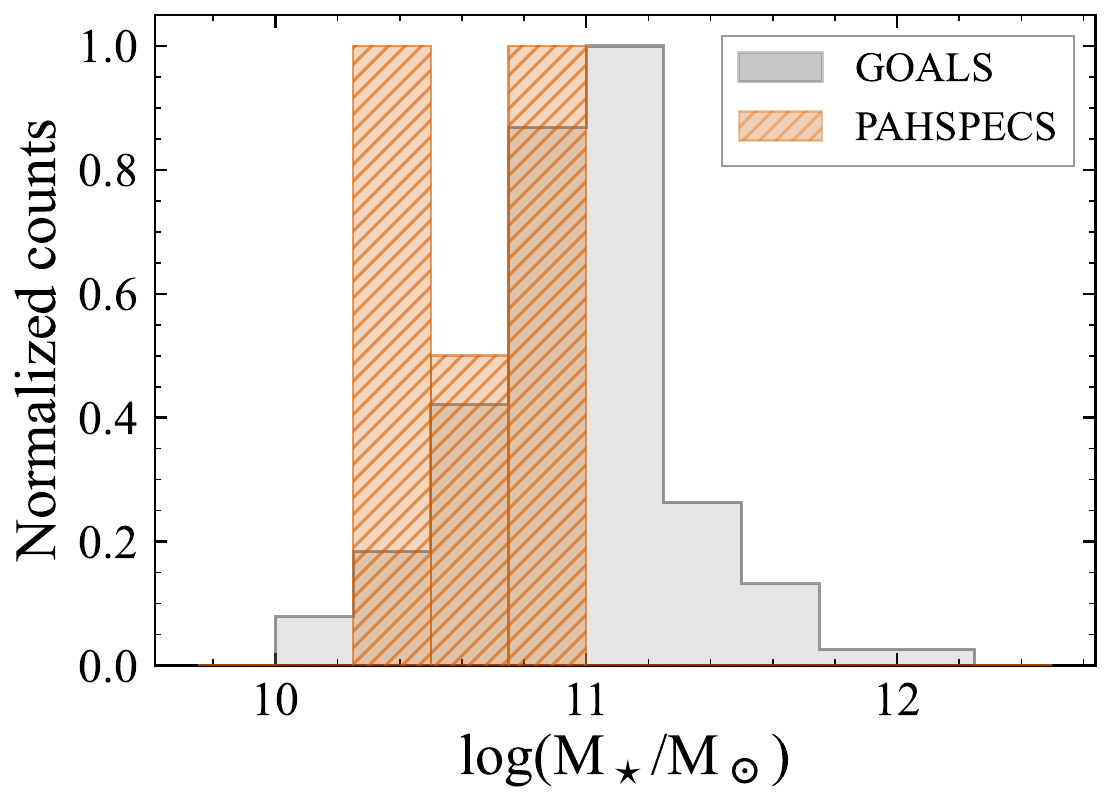}
    \end{subfigure}
    \hfill
    \begin{subfigure}[b]{0.24\textwidth}
        \includegraphics[width=\textwidth]{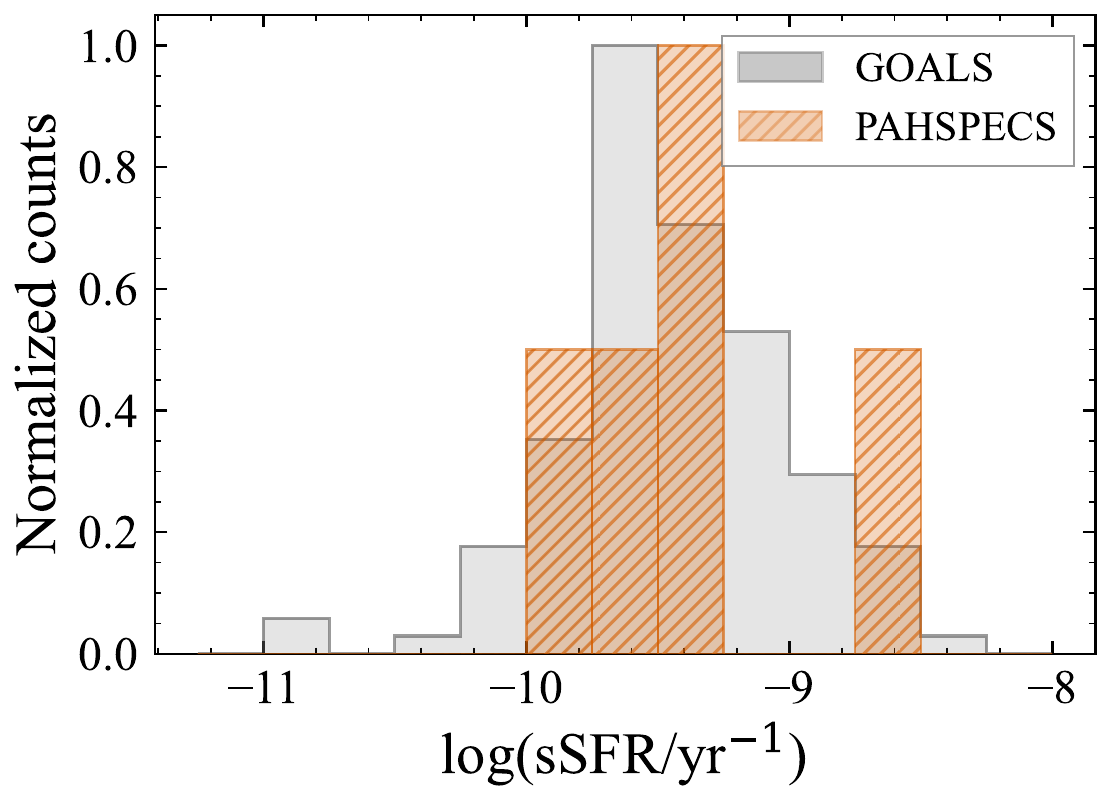}
    \end{subfigure}
    \caption{Distribution of global galaxy properties for the PAHSPECS sample compared to the local GOALS galaxies. From left to right: infrared luminosity $\rm \log(\rm L_{\rm IR}/ \rm L_{\odot}$), star-formation rate $\rm \log (\mathrm{\rm SFR/\rm M_{\odot} yr^{-1}}$), stellar mass $\rm \log (\rm M_\star/M_{\odot}$), and specific SFR $\rm \log (\mathrm{\rm sSFR/yr^{-1}}$). The gray histograms represent the GOALS sample, while the hatched orange histograms indicate the PAHSPECS galaxies. The PAHSPECS galaxies occupy the low-to-intermediate luminosity and mass regime of the GOALS distribution, with comparable sSFRs.}
    \label{fig:histograms}
\end{figure*}

In Stage 3, the \texttt{Spec3Pipeline} was used to construct and combine 3D IFU cubes from dithered exposures using association files. Outlier rejection and bad-pixel replacement were enabled, with a conservative threshold adopted for identifying and masking deviant pixels. The cube-building step was performed in the IFU-aligned coordinate system, so that the input exposures are aligned in detector coordinates. The resulting Stage 3 products (\texttt{*_s3d.fits}) are flux-calibrated and fully aligned 3D data cubes, and consists of 12 data cubes, corresponding to each channel (1,2,3 and 4)
and the respective sub-bands (SHORT, MEDIUM and LONG configurations), providing a continuous wavelength coverage between
4.9–27.9~$\mu m$.  

We noticed that the fully calibrated data cubes still contained low-level artifacts that appear as stripes aligned with the detector axes. To correct for these artifacts we apply a destriping procedure in which, after the source region was masked, the median signal was computed for each row of the cube and subtracted from all pixels in the corresponding row. This consistently reduces the noise level, as quantified by measuring a lower standard deviation in the destriped spectra compared to their non-destriped version. As an additional validation, we performed synthetic photometry on the reduced spectra using the F1500W ($\lambda_{\rm rest}$\,=\,7.7~$\mu m$) and F770W ($\lambda_{\rm rest}$\,=\,3.3~$\mu m$) filters. The fluxes derived from the destriped cubes show better agreement with the imaging measurements than those obtained from the non-destriped data.

\section{Spectral extraction and fitting}
\label{spectral extraction}

From the MIRI/MRS data cubes described in Sect.~\ref{sample and JWST reduction}, we extracted one-dimensional spectra using a wavelength-dependent aperture centered on each source. Due to the faintness of the sources in the MRS cubes, the extraction aperture was determined empirically from imaging photometry obtained as part of the Systematic Mid-infrared Instrument Legacy Extragalactic (SMILES\footnote{\href{https://archive.stsci.edu/hlsp/smiles}{https://archive.stsci.edu/hlsp/smiles}}; \citealt{Alberts24, Rieke24}) survey, using the MIRI filters F560W, F770W, F1000W, F1280W, F1500W, F1800W, F2100W and F2550W. For each band, the aperture radius was selected from the source curve of growth to maximize the signal-to-noise ratio (SNR), typically enclosing $\sim$60\% of the total flux and requiring an aperture correction of $\sim$1.7. We then fitted a linear function to these reference radii as function of wavelength. Specifically, we parameterized the aperture radius as $r[''] = m \times \lambda[\mu m] + b['']$, where $m$ and $b$ are the slope and intercept, respectively. The resulting $(m,b)$ values are (0.0102, 0.5822) for ASPECS-6, (0.0072, 0.3641) for ASPECS-11, (0.0129, 0.2339) for ASPECS-14, (0.0117, 0.5419) for ASPECS-15, and (0.0106, 0.3714) for ASPECS-C20.  

To model the MIR spectral features, we employed a modified version of the Continuum And Feature Extraction code \texttt{CAFE}\footnote{\href{https://github.com/GOALS-survey/CAFE}{https://github.com/GOALS-survey/CAFE}} originally developed by \citet{Marshall07} for Spitzer/IRS and recently adapted for JWST \citep{CAFE25}. The spectra are simultaneously decomposed into multiple components, including emission from stellar populations (the interstellar radiation field), dust continuum emission, PAH features, atomic and molecular gas emission lines, and dust attenuation. PAH features are modeled using Drude profiles, with individual bands represented either by single components or by the sum of multiple sub-components. The 3.3 and 6.2~$\mu$m bands are each described by a single Drude component, while the 7.7 and 11.3~$\mu$m complexes are decomposed into multiple sub-components. In particular, the 7.7~$\mu$m feature includes the 7.42, 7.60, and 7.85~$\mu$m components, while the 11.3~$\mu$m feature includes the 11.05, 11.23, and 11.33~$\mu$m components. For these complex features, the total flux is obtained by summing the contributions of the individual sub-components. We note that atomic and molecular lines are not included in our fits and will be explored in more detail in future work. We adopt a foreground screen geometry for the attenuation of the cold, cool, warm, and hot dust continuum components, while the PAH emission is treated with a mixed geometry in which the PAH molecules are mixed with the surrounding dust. The extinction curve, primarily constrained by the strength of the 9.7~$\mu$m silicate absorption feature ($\tau_{9.7}$), is used to correct the PAH fluxes. When fitting only the MIRI/MRS spectra, we obtain relatively high optical depths on the order of $\tau_{9.7} {\sim}$3--4. These are values typically associated with highly obscured systems such as local LIRGs \citep{Sabrina14}. To improve the quality and physical consistency of the fits, we incorporate ancillary JWST/NIRCam photometry from the JADES catalogue in the F335M, F356W, F410M, F430M, F444W, F460M, and F480M filters, together with Herschel/PACS photometry at 100 and 160~$\mu$m and ALMA 1~mm continuum measurements (see Fig.~\ref{fig: CAFE fit aspecs6} and Appendix~\ref{appendix with CAFE fits}). This results in lower optical depths, with $\tau_{9.7}$ ranging from 0.3 to 0.9, except for ASPECS-15, the only AGN in the sample, which reaches $\tau_{9.7}$ = 1.9 (see Table~\ref{tab:pahspecs_sample_props}). Table~\ref{tab:pah_fluxes} presents the PAH luminosities both corrected and uncorrected for extinction together with their uncertainties estimated from the fits.

Upper limits were derived for sources without a significant detection of the 3.3~$\mu$m PAH feature (ASPECS-11, ASPECS-15, and ASPECS-C20). We estimated the noise empirically from the data cubes using off-source extractions, since the uncertainties derived from the MRS variance cubes underestimate the true noise. For ASPECS-11 and ASPECS-C20, where source-free areas are available within the field of view, we extracted off-source spectra using the same wavelength-dependent apertures adopted for the source spectra. One sigma upper limits are then estimated by computing the root mean square (RMS) over a wavelength interval of $\pm$~2\,FWHM (corresponding to $\Delta\lambda{\sim}$0.1\,$\mu m$, as defined in \citet{Draine07}) around the wavelength of the feature, $\lambda_{\rm rest}$ = 3.29~$\mu$m. We note that the off-source spectra show low-level spectral wiggles, indicating that the noise is partially correlated across wavelength channels. To account for this, we multiply the uncertainty by the square root of the number of correlated spectral elements across the integration window. This provides a more conservative estimate of the effective $1\sigma$ uncertainty. For ASPECS-15, where the field of view is crowded and off-source extraction is not feasible, we instead derive the noise directly from the observed spectrum. We adopt the same wavelength interval used for ASPECS-11 and ASPECS-C20 and compute a local continuum from two spectral windows adjacent to the feature. We then subtract the continuum, measure the RMS from the residuals, and estimate the $1\sigma$ upper limit by quadratically integrating the RMS over the feature extent as defined before. 1$\sigma$ upper limits for the 3.3 $\mu m$ feature are reported in Table~\ref{tab:pah_fluxes}.

We further performed spectral energy distribution (SED) fitting of the photometry with the \texttt{Prospector} code \citep{Prospector}, using publicly available photometry from rest-frame UV to sub-mm, to derive estimates of the stellar mass ($M_{\star,\mathrm{SED}}$) and SFR$_{\mathrm{SED}}$. For a full description of the SED-fitting procedure we refer the reader to \citet{Shivaei26} and \citet{Boogaard26}. We also derived the effective radius at 15~$\mu$m, $r_{e}^{15\mu\mathrm{m}}$, from the simultaneous, deep MIRI 15~$\mu$m imaging from our program (see \citet{Shivaei26}), given its high SNR. The corresponding values are reported in Table~\ref{tab:pahspecs_sample_props}.

\section{Results}
\label{results}

Figure~\ref{fig:histograms} presents the distribution of $\rm L_{\rm IR}$, SFR, and $M_\star$ of the PAHSPECS galaxies with respect to the nearby reference sample of dusty, luminous galaxies from GOALS. Although the two samples are not drawn from the same parent galaxy population (PAHSPECS traces main-sequence star-forming galaxies at $z\sim1$, whereas GOALS consists of nearby starburst systems), the PAHSPECS galaxies overlap with the low-to-intermediate luminosity and stellar-mass range of the local LIRG sample and exhibit comparable SFRs. The wider parameter range covered by GOALS provides a local reference for examining possible scaling relations and identifying offsets in the PAH properties of the PAHSPECS galaxies. For GOALS, we use the AKARI/IRC 2.5–5 $\mu$m and Spitzer/IRS 5–38 $\mu$m spectra from \citet{Sabrina13, Sabrina14, Inami18}, which we re-fit with the new version of \texttt{CAFE} for consistency. We further adopt $f_{\mathrm{AGN}}$, defined as the fractional contribution of the AGN to the MIR spectrum, from \citet{Diaz-Santos17}, L$_{IR}$ from \citet{Armus09}, M$_{\star}$ and SFR from \citet{Howell10}.

\subsection{PAH Ratio Diagnostics} 

\begin{figure*}[p]
    \centering
    \begin{subfigure}[b]{0.7\textwidth}
        \centering
        \includegraphics[width=\textwidth]{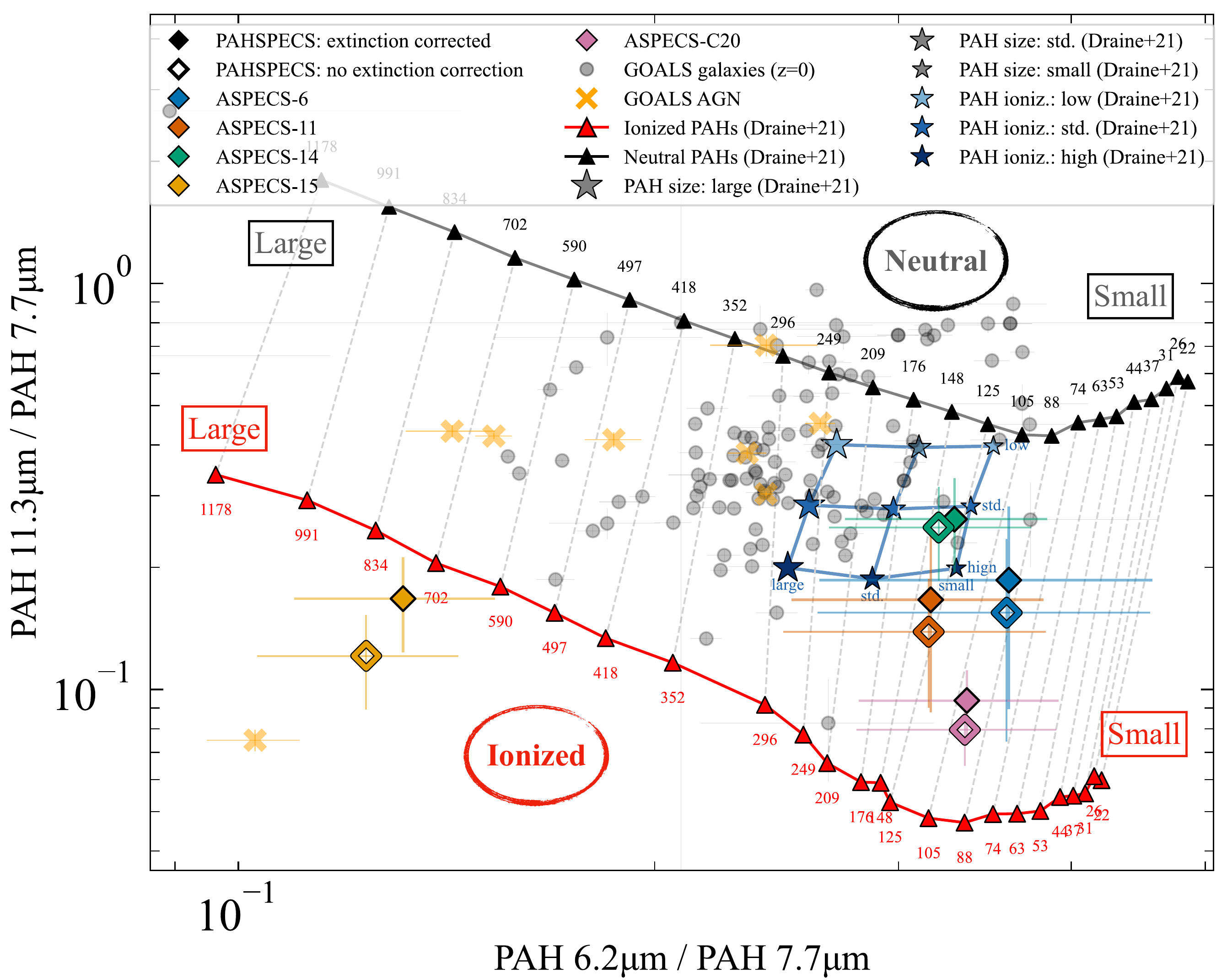}
    \end{subfigure}

    \vspace{6pt}

    \begin{subfigure}[b]{0.7\textwidth}
        \centering
        \includegraphics[width=\textwidth]{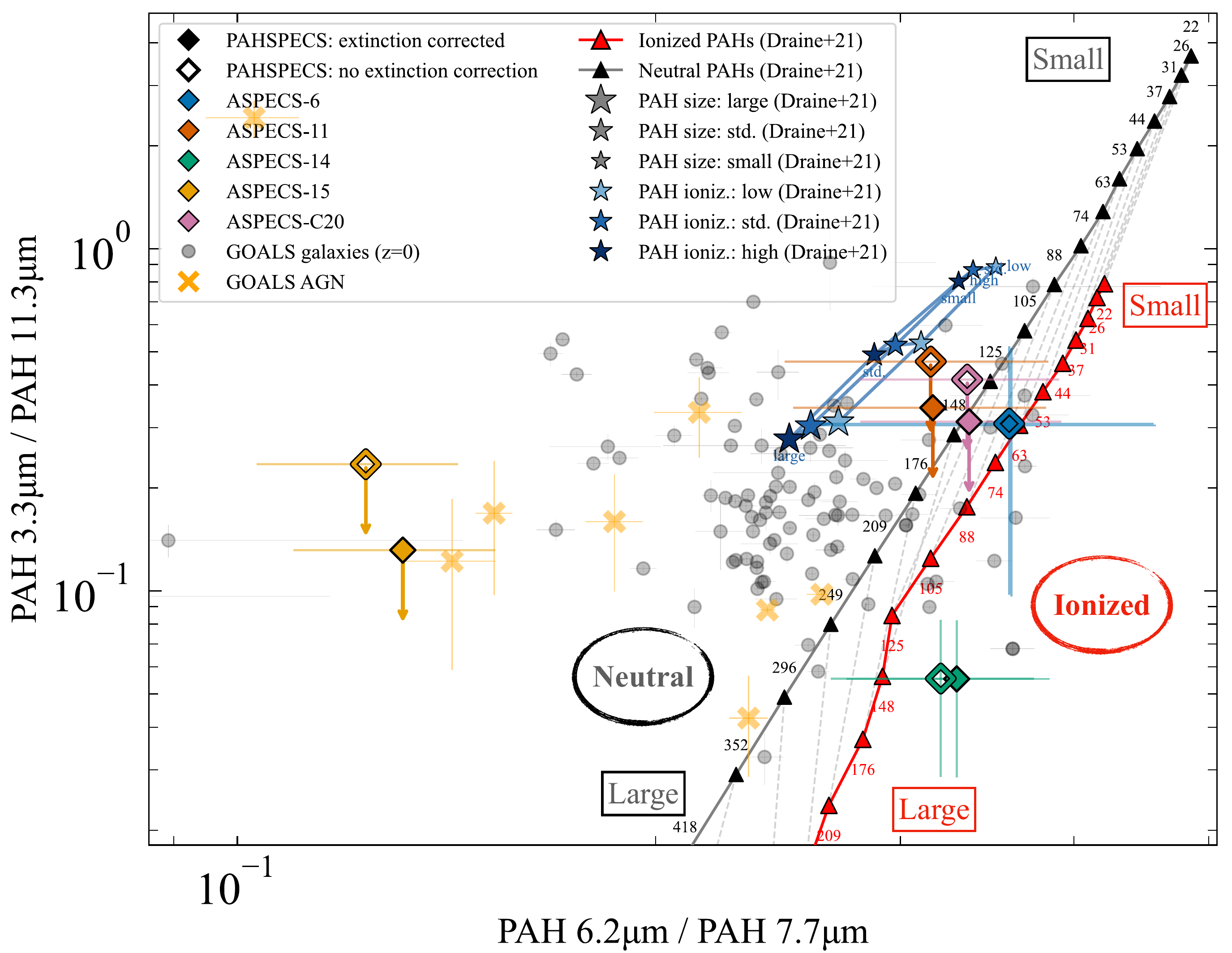}
    \end{subfigure}

    \caption{PAH band-ratio diagrams comparing the PAHSPECS galaxies with the local LIRGs: PAH~11.3/7.7 versus PAH~6.2/7.7 (\textit{top}), PAH~3.3/11.3 versus PAH~6.2/7.7 (\textit{middle}), and PAH~11.3/7.7 versus PAH~3.3/11.3 (\textit{bottom}; shown on the following page). The PAHSPECS galaxies are indicated by colored symbols, with filled and open markers corresponding to extinction-corrected and observed measurements, respectively and $3\sigma$ upper limits are reported. GOALS galaxies are shown as grey circles, with AGN-dominated sources marked with crosses. For reference, we overplot the model tracks from \citet{Draine21} for individual neutral and ionized PAHs (black and red triangles, respectively), with the labels indicating the number of carbon atoms, $N_{\rm C}$, together with the blue grid representing PAH populations with different size and distributions.}
    \label{fig:pah_ratios}
\end{figure*}

\begin{figure*}[t]
    \ContinuedFloat
    \centering
    \begin{subfigure}[b]{0.7\textwidth}
        \centering
        \includegraphics[width=\textwidth]{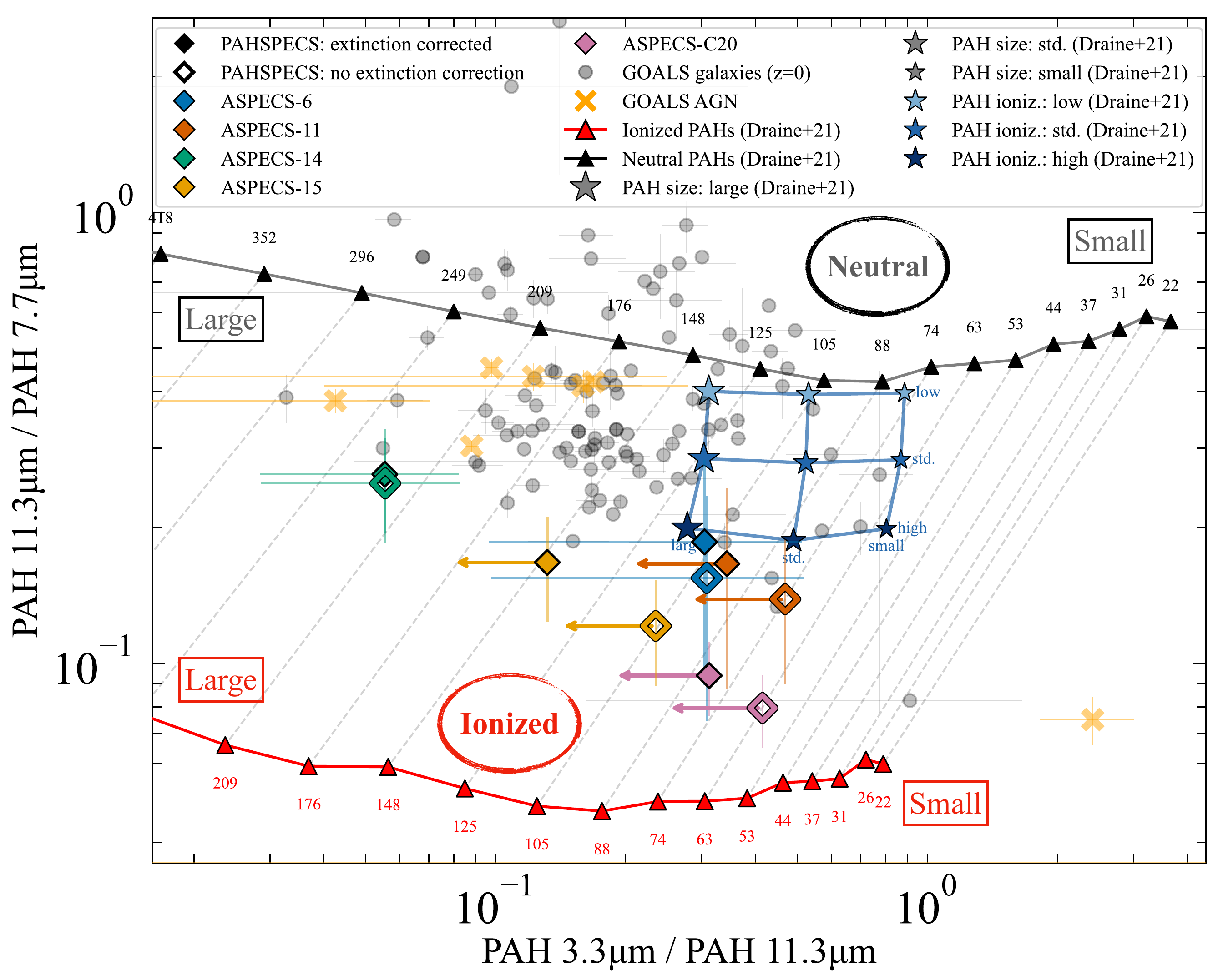}
    \end{subfigure}

    \caption{Continued. PAH~11.3/7.7 versus PAH~3.3/11.3 (\textit{bottom}).}
\end{figure*}


Figure~\ref{fig:pah_ratios} compares the PAHSPECS galaxies with the nearby starburst sample across three PAH band-ratio planes: 11.3/7.7 versus 6.2/7.7 (\textit{top}), 3.3/11.3 versus 6.2/7.7 (\textit{middle}), and 11.3/7.7 versus 3.3/11.3 (\textit{bottom}). The 11.3/7.7 PAH ratio primarily traces the PAH ionization state, with lower values indicating a larger contribution from ionized grains \citep[e.g.,][]{Sabrina14, Maragkoudakis18, Lai22, Donnelly24, Donnan26}, while the 6.2/7.7 \citep[e.g.,][]{Pereira10, Sabrina14, Garcia-Bernete22, Donnan26} and 3.3/11.3 \citep[e.g.,][]{Maragkoudakis20, Lai23, McKinney25, Donnan26} ratios are commonly used as diagnostics of the characteristic PAH size distribution, with higher 3.3/11.3 and 6.2/7.7 PAH ratios associated with smaller grains. However, we note that these ratios are also sensitive to the radiation field, as variations in the energy distribution of absorbed photons can modify the relative strength of short- and long-wavelength PAH features even at fixed size and charge \citep{Draine21}. 

\begin{figure*}[] 
    \centering
    \includegraphics[width=\textwidth]{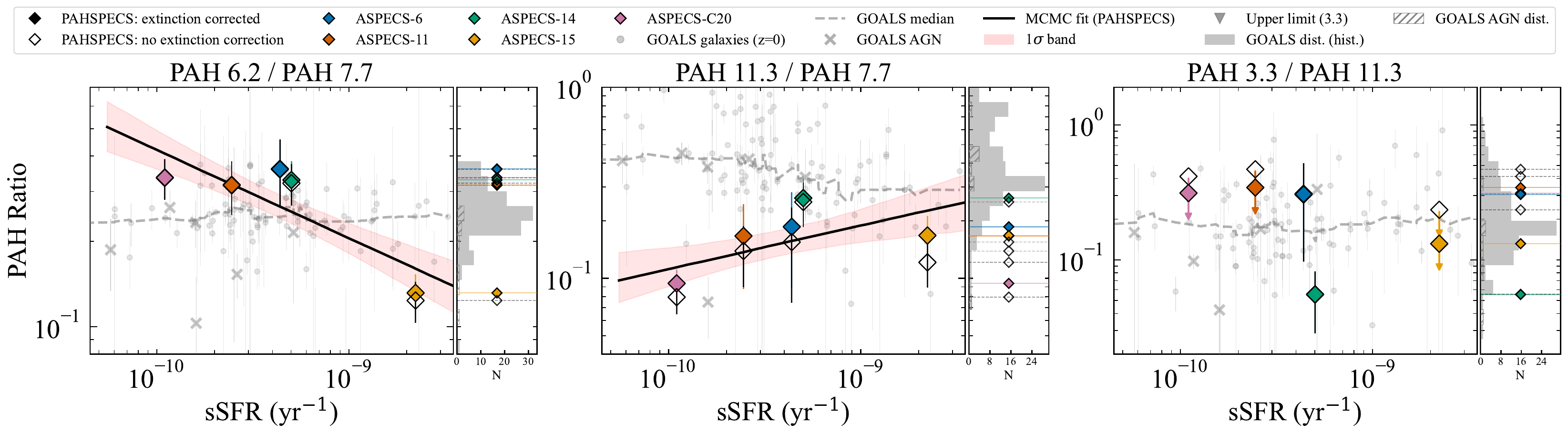}
    \caption{PAH band ratios as a function of sSFR: PAH~6.2/7.7 (\textit{left}), PAH~11.3/7.7 (\textit{middle}), and PAH~3.3/11.3 (\textit{right}). Symbols follow the convention of Fig.~\ref{fig:pah_ratios}. Dashed lines show the running medians of the GOALS sample. For PAH~6.2/7.7 and PAH~11.3/7.7, solid lines and shaded regions indicate the MCMC best-fit relations for the PAHSPECS galaxies and their $1\sigma$ uncertainties, respectively. The marginal panels show the corresponding GOALS distributions and the positions of the PAHSPECS galaxies.}
    \label{fig:ssfr PAH ratios}
\end{figure*}

In the 11.3/7.7 versus 6.2/7.7 plane (Fig.~\ref{fig:pah_ratios}, \textit{top} panel), most of the PAHSPECS galaxies are offset toward higher 6.2/7.7 values (PAH~6.2/7.7~$\sim$~0.35) relative to the bulk of the local LIRG population. While the local systems span a broad range in 11.3/7.7, extending towards more neutral-dominated values, most of the PAHSPECS sources cluster around 0.1${\leq}$PAH~11.3/7.7${\leq}$0.3, in an area intermediate between the two model tracks of \citet{Draine21}, with a preference toward the more ionized one.
Notably, ASPECS-15, the only identified AGN host in the PAHSPECS sample, lies beyond the ionized-PAH track of \citet{Draine21}, although it remains compatible with the model within the uncertainties. It exhibits both the lowest PAH~6.2/7.7 ($\sim0.12$) and one of the lowest PAH~11.3/7.7 ($\sim0.1$) ratios among the PAHSPECS galaxies. Moreover, both ratios are lower than those typically observed in the local AGN-dominated systems, indicating that ASPECS-15 extends toward a regime characterized by a reduced relative contribution from small PAHs and a stronger contribution from ionized PAHs.We note that ASPECS-15 is identified as an X-ray AGN, whereas AGN in the GOALS sample are selected based on their MIR properties; while these selection methods may probe different AGN populations, the comparison remains informative in the context of their MIR spectral properties. 
 
The PAH~3.3/11.3 ratio, shown in the middle and bottom panels of Fig.~\ref{fig:pah_ratios}, is detected only for ASPECS-6 and ASPECS-14, which lie toward opposite ends of the local LIRG distribution. For ASPECS-11, ASPECS-15, and ASPECS-C20, only $3\sigma$ upper limits are available (see Sect.~\ref{spectral extraction}). Consistent with our findings, the 3.3~$\mu$m PAH feature is neither robustly detected in ASPECS-11, ASPECS-15, and ASPECS-C20 in the resolved study of \citet{Donnan26}. Given the limited number of detections, no firm conclusion can be drawn regarding systematic differences in PAH~3.3/11.3 between the PAHSPECS and the local sample. Nevertheless, the detections together with the upper limits are compatible with a scenario in which the neutral PAH population in the PAHSPECS galaxies is weighted toward larger grains relative to local systems; this possibility is discussed further in Sect.~\ref{PAH ratio discussion} and Sect.~\ref{3.3 Discussion}.

\subsection{PAH Ratios as a Function of sSFR}
Figure~\ref{fig:ssfr PAH ratios} presents the PAH~6.2/7.7, PAH~11.3/7.7, and PAH~3.3/11.3 ratios as a function of sSFR in the left, middle, and right panels, respectively. These diagrams probe whether PAH size and ionization diagnostics vary with the star-forming conditions of the galaxies.
The PAH~6.2/7.7 and PAH~11.3/7.7 ratios display contrasting trends with sSFR in the local and cosmic-noon samples. In nearby galaxies, PAH~6.2/7.7 remains approximately constant across the sampled sSFR range, whereas it decreases in the PAHSPECS galaxies, with a slope of $\sim-0.3$. Conversely, PAH~11.3/7.7 increases with sSFR in the PAHSPECS sample, with a slope of $\sim+0.2$, while showing a slight decline in the nearby galaxies.
For the PAH~3.3/11.3 ratio, the two detected PAHSPECS sources lie toward opposite ends of the nearby LIRG distribution, while the remaining three sources are constrained only by $3\sigma$ upper limits. Given the limited number of significant detections of the 3.3~$\mu$m feature (two out of five sources), we do not attempt to fit a trend with sSFR.

\subsection{PAH-to-IR Ratios as a Function of Global Galaxy Properties}

\begin{figure}[h] 
    \centering
    \includegraphics[width=\columnwidth]{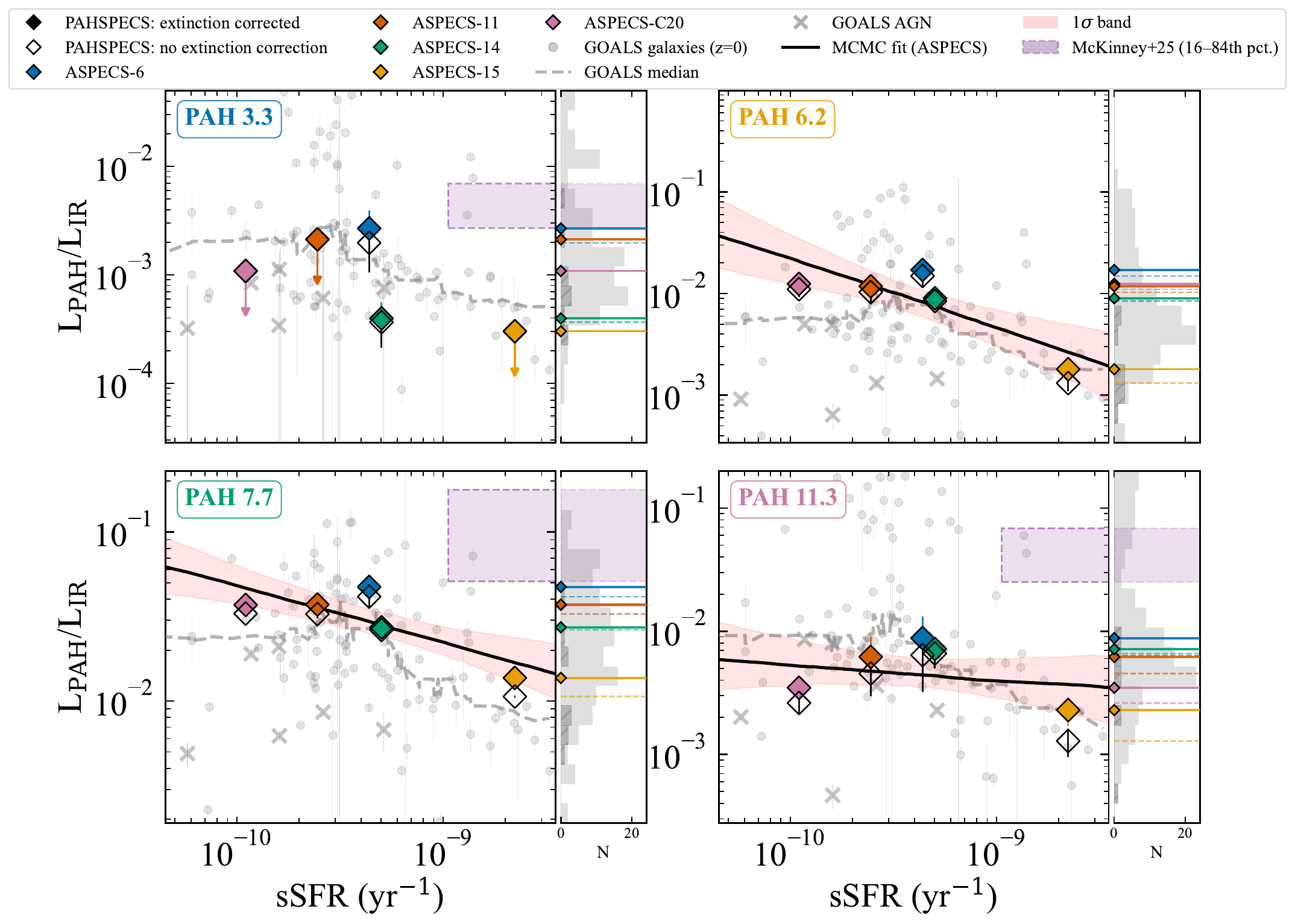}
    \caption{$L_{\rm PAH}/L_{\rm IR}$ as a function of sSFR for the 3.3, 6.2, 7.7, and 11.3~$\mu$m PAH features (\textit{upper left}, \textit{upper right}, \textit{lower left}, and \textit{lower right}, respectively). Symbols follow the convention of Fig.~\ref{fig:pah_ratios}. The purple shaded region indicates the 16th--84th percentile range of the cosmic-noon sample from \citet{McKinney25}.}
    \label{fig: ssfr vs LPAH LIR}
\end{figure}

Figure~\ref{fig: ssfr vs LPAH LIR} presents $L_{\rm PAH}/L_{\rm IR}$ as a function of sSFR for the PAH features at 3.3, 6.2, 7.7, and 11.3~$\mu$m, probing their relative contribution to the total infrared emission across different levels of star-formation activity. The PAHSPECS measurements of the 11.3~$\mu$m~PAH are broadly consistent with the local median trend, while those of 6.2 and 7.7~$\mu$m lie slightly above the nearby-sample medians. For the 3.3~$\mu$m~PAH, ASPECS-6 and ASPECS-14 fall within the local distribution, while the remaining sources are constrained only by upper limits.

\subsection{PAH 7.7–SFR Scaling and the $\rm L_{7.7}/\rm L_{\rm IR}$--6.2~EQW diagnostic}

To investigate whether the PAH emission in the PAHSPECS galaxies follows the same relation with star formation as in nearby IR-luminous systems, we first compare the PAH~7.7~$\mu$m luminosity with the SFR. We then examine $L_{7.7}/L_{\rm IR}$ as a function of the 6.2~$\mu$m PAH equivalent width, 6.2~EQW, to assess the relative contribution of PAH emission and the possible influence of an AGN-dominated MIR continuum.
Figure~\ref{fig:SFR_7.7} shows the 7.7~$\mu$m PAH luminosity as a function of SFR for both the cosmic noon and local galaxy samples, together with the 7.7~PAH–SFR scaling relations from \citet{Shipley16} and \citet{Kim24} shown for comparison.
The PAHSPECS galaxies lie within the locus defined by the nearby IR-luminous sample across the full range of SFRs covered by our data, and fall along the same general trend traced by the nearby galaxies.  

Figure~\ref{fig:6.2_EW} presents the ratio $\rm L_{7.7}/\rm L_{\rm IR}$ as a function of the 6.2~EQW. We estimated the uncertainties on the EW by propagating the measured flux uncertainties. The continuum level was measured  around the 6.2~$\mu$m feature, and its uncertainty was estimated from the RMS of the off-source spectrum extracted from an empty region of the cube. 
In this plane, the local AGN, shown as crosses, preferentially occupy the region of lower 6.2~EQW and $L_{7.7}/L_{\rm IR}$, while pure starbursting galaxies extend toward higher 6.2~EQW and $L_{7.7}/L_{\rm IR}$. The galaxies in PAHSPECS are preferentially located toward the high-6.2~EQW side of the plot (0.5 $\mu m \leq $6.2~EQW $\leq$ 1.4 $\mu m$) and span a relatively narrow range in $\rm L_{7.7}/L_{\rm IR}$ (0.015~$\leq \rm L_{7.7}/L_{\rm IR}~\leq$~0.06) compared to the full GOALS distribution. The AGN-hosting source in the PAHSPECS sample, ASPECS-15, has the lowest 6.2~EQW among the cosmic noon galaxies (${\sim}$0.5 $\mu m$) and the lowest $L_{7.7}/L_{\rm IR}$~($\sim$~0.015).

\section{Discussion}
\label{discussion}

\subsection{PAH ratio diagnostics in the context of local and cosmic-noon galaxies}
\label{PAH ratio discussion}

\begin{figure}[t]
    \centering
    \includegraphics[width=\linewidth]{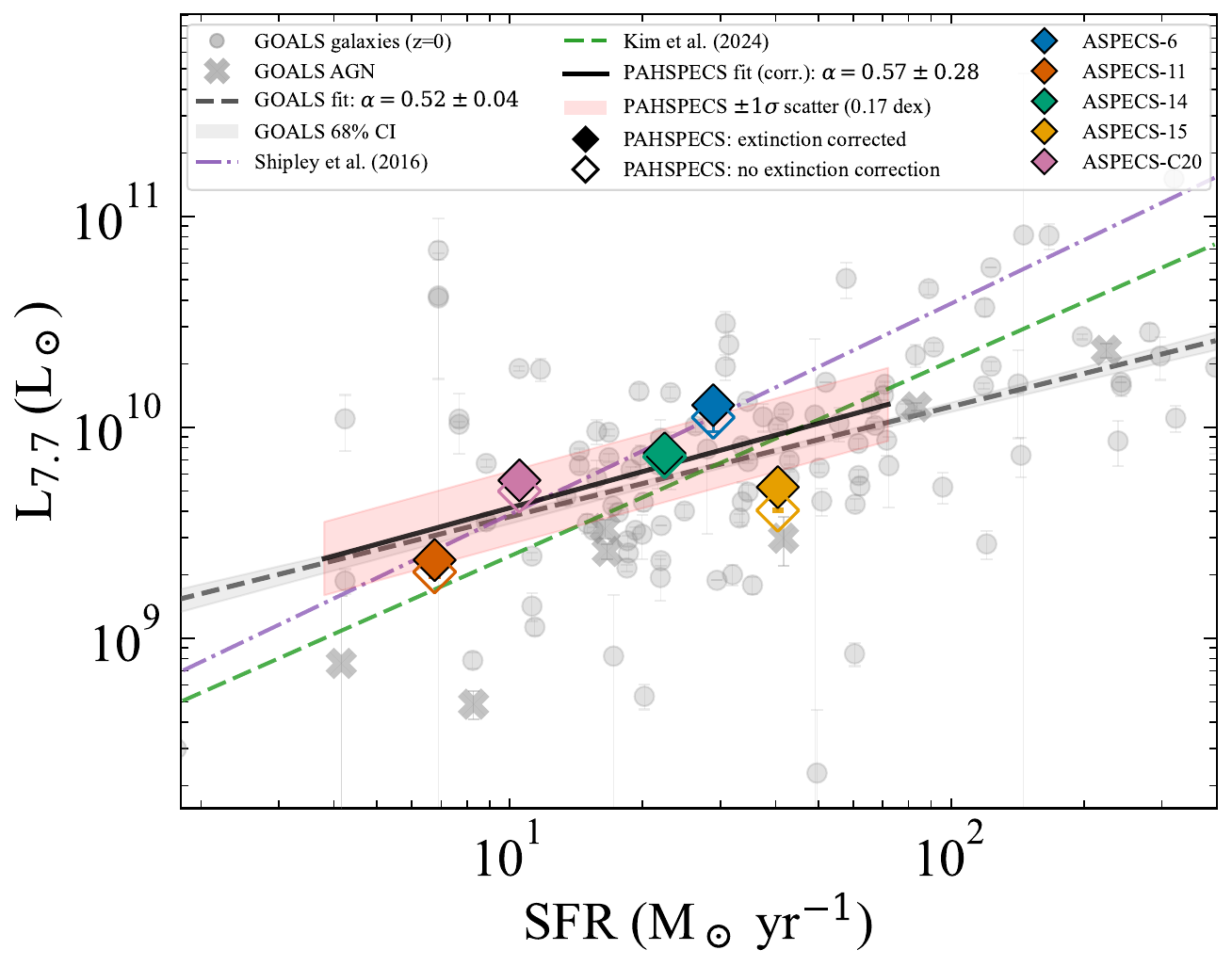}
    \caption{7.7~$\mu$m PAH luminosity as a function of SFR. Symbols follow the convention of Fig.~\ref{fig:ssfr PAH ratios}. A best-fit relation to the local sample is overplotted together with its 68\% confidence interval. For reference, we additionally include the 7.7~PAH–SFR relations from \citet{Shipley16} and \citet{Kim24}.}
    \label{fig:SFR_7.7}
\end{figure}

\begin{figure}[t]
    \centering
    \includegraphics[width=\linewidth]{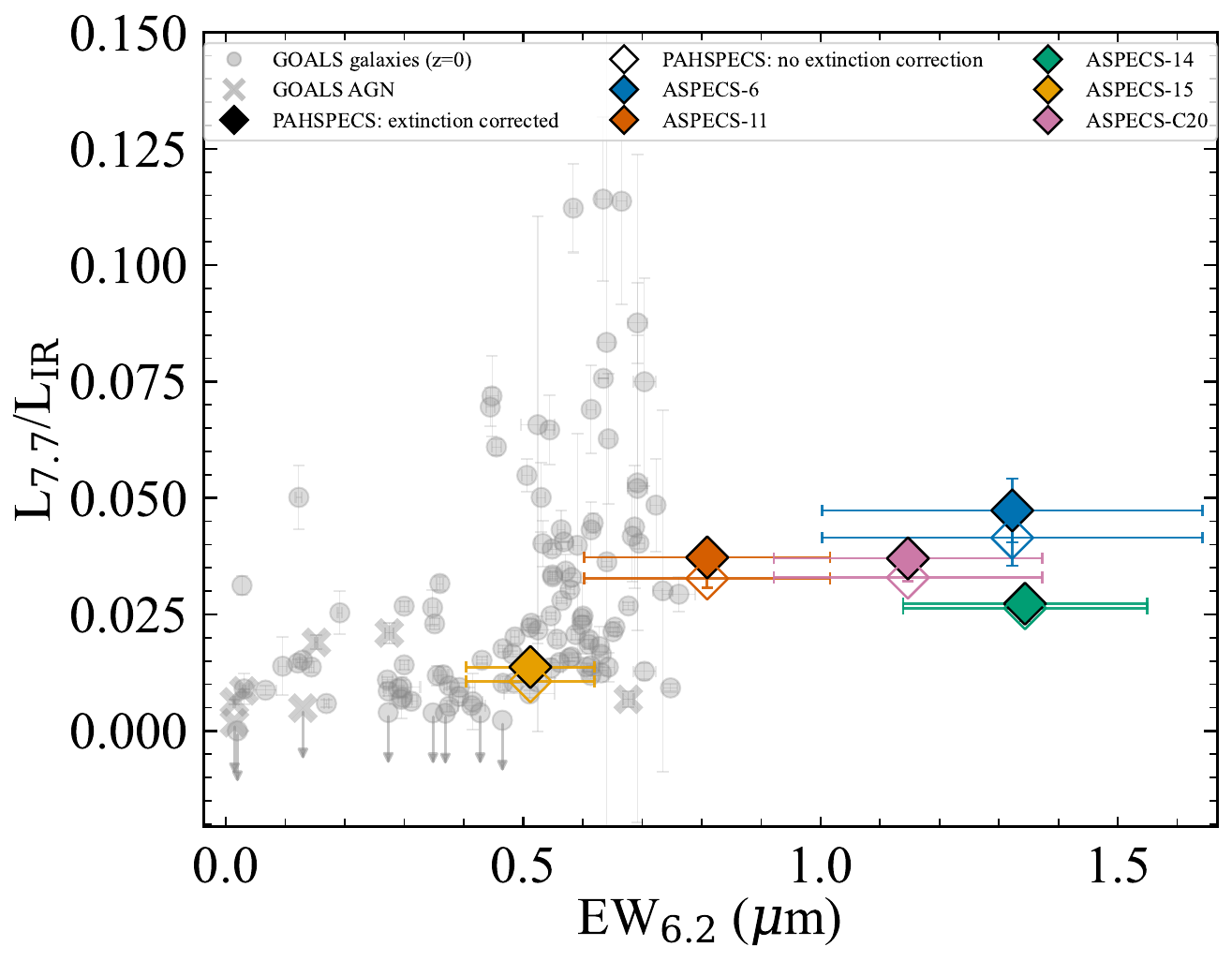}
    \caption{$\rm L_{7.7}/\rm L_{\rm IR}$ versus the 6.2~$\mu$m equivalent width (6.2~EQW). Symbols follow the same convention as in Fig.~\ref{fig:ssfr PAH ratios}.}
    \label{fig:6.2_EW}
\end{figure}

High-z galaxies are characterized by higher ionization parameters at a given M$_{\star}$ and SFR \citep{Kaasinen18, Sanders20, Shen25} and lower gas-phase metallicities at fixed M$_{\star}$ \citep[e.g.,][]{Shivaei17, Henry21, Maiolino19, Sanders21} compared to nearby, luminous dusty galaxies ---conditions that are known to influence PAH emission. As a consequence, we would expect systematic differences in the PAH properties (ionization state and grain-size distribution) between cosmic noon and local galaxies.


To investigate whether these different physical conditions are reflected in the PAH populations, we first consider the PAH~6.2/7.7 and PAH~11.3/7.7 ratios, which provide complementary information on the characteristic PAH size distribution and ionization state. Both ratios show systematic differences relative to the nearby galaxies: in Fig.~\ref{fig:pah_ratios} four out of five PAHSPECS sources exhibit higher PAH~6.2/7.7 and lower PAH~11.3/7.7 ratios than the bulk of the nearby LIRGs, suggesting that the ionized PAH component in the PAHSPECS galaxies may be weighted toward smaller grains. In terms of the PAH~6.2/7.7 ratio, this result is in line with spatially resolved \textit{Spitzer}/IRS spectroscopy of nearby H{\sc II} galaxies, Seyfert galaxies, and LIRGs by \citet{Garcia-Bernete22}, who find that normal star-forming regions display slightly higher PAH~6.2/7.7 ratios than local LIRGs, clustering around PAH~6.2/7.7${\sim}0.3$ rather than ${\sim} 0.25$. Although those measurements are spatially resolved, they indicate that, at least with respect to the PAH~6.2/7.7 ratio, the PAHSPECS locus is not unique to extreme local systems, but overlaps with that of normal star-forming environments on smaller scales.
In contrast, ASPECS-15, the only identified AGN host in the PAHSPECS sample, lies outside the PAH~6.2/7.7--PAH~11.3/7.7 locus typically occupied by local AGN-dominated systems, exhibiting lower values of both ratios. This offset may point to differences in the PAH populations and/or ISM conditions of AGN hosts at cosmic noon relative to their local counterparts.
The PAH~3.3/11.3 ratio provides a complementary, although less  constraining, diagnostic of the PAH size distribution in the PAHSPECS sample, owing to the limited number of detections of the 3.3~$\mu$m feature. Nevertheless, our measurements remain compatible with the tendency in which the neutral PAH population is biased toward larger grains than in nearby systems. This scenario is also suggested by the spatially resolved analysis of \citet{Donnan26}, who identify regions in ASPECS-6 with high PAH~6.2/7.7 and low PAH~3.3/11.3 ratios, and by \citet{McKinney25}, who report PAH~11.3/3.3 ratios in cosmic-noon starburst systems that are, on average, ${\sim}3$ times higher than those of local LIRGs. The limited constraints on the 3.3~$\mu$m feature in our sample therefore neither confirm nor rule out this scenario; the possible origin of the non-detections is discussed further in Sect.~\ref{3.3 Discussion}.

\begin{figure}[] 
    \centering
    \includegraphics[width=\columnwidth]{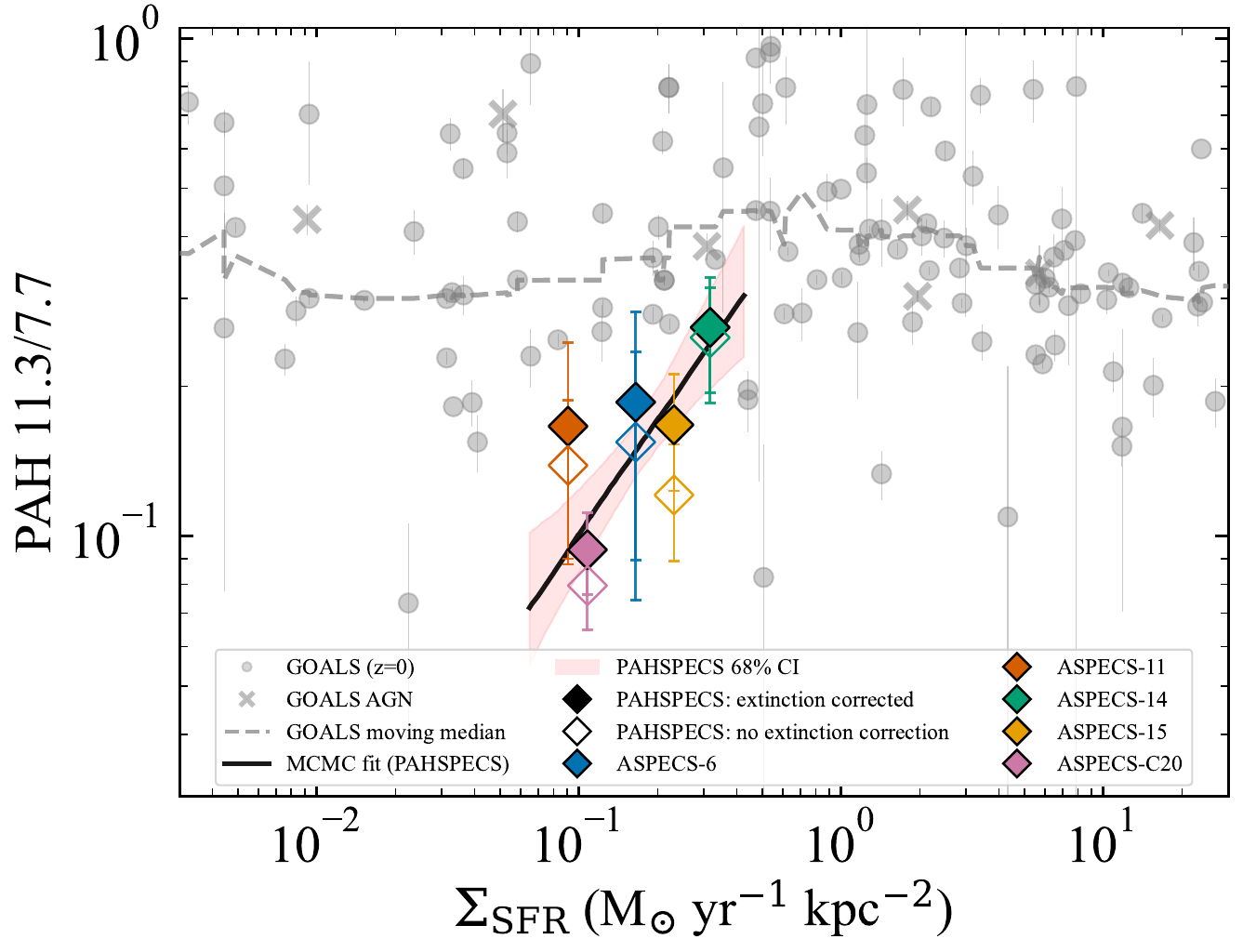}
    \caption{11.3/7.7 as function of the SFR surface density, $\Sigma_{\rm SFR}$. Symbols follow the same convention as in Fig.~\ref{fig:ssfr PAH ratios}.}
    \label{fig: SFR surface density vs 11/7}
\end{figure}

\subsection{Interpreting the Variations in PAH Band Ratios}
\label{PAH vs sSFR}
To explore whether global star formation activity regulates the PAH properties in the PAHSPECS galaxies, we examine the behavior of 11.3/7.7, 6.2/7.7, and 3.3/11.3 as a function of sSFR (Fig.~\ref{fig:ssfr PAH ratios}). We see a positive trend between 11.3/7.7 and sSFR within the PAHSPECS sample (\textit{middle} panel) which approaches the nearby starburst systems only at high sSFR. However, given the small sample size, we regard the correlation only as tentative. One potential interpretation of the increasing 11.3/7.7 with sSFR might be related to variations in radiation field intensity and hardness, which regulate the balance between neutral and ionized PAH components \citep{Draine01a, Draine01b, Weingartner01, Draine21}. \citet{Shen25} show that [O\,{\sc III}]$_{\lambda\lambda 4959,5007}$/[O\,{\sc II}]$_{\lambda\lambda 3726,3729}$, a tracer of the ionization parameter, increases with sSFR in star-forming galaxies at $1{\lesssim}z{\lesssim}3$. If higher sSFR systems indeed host more intense radiation fields, a plausible interpretation is that smaller and more highly ionized PAHs are preferentially destroyed under these conditions, while larger and more neutral grains are comparatively resilient \citep[e.g.][]{Leach86, Leger89, Voit92, Allain95, Jochims99, Zhen15}. Such selective processing would naturally enhance the relative strength of the 11.3 $\mu m$ over the 7.7 $\mu m$ feature, leading to an increase in 11.3/7.7 toward higher sSFR. Indeed, Fig.~\ref{fig: ssfr vs LPAH LIR} shows a decline in $L_{\rm PAH}/L_{\rm IR}$ with sSFR for the 6.2 and 7.7~$\mu$m features, while the 11.3~$\mu$m feature remains relatively flat. This is consistent with the selective destruction of smaller, more ionized PAHs, favoring the survival of larger and more neutral grains. This is fully consistent with the resolved study of ASPECS-6 in \citet{Donnan26}, which finds that the 11.3/7.7 PAH ratio increases with both the hardness of the radiation field and the sSFR of the source.
Fig.~\ref{fig:ssfr PAH ratios} also examines the 6.2/7.7 ratio (\textit{left} panel) as a function of sSFR. The 6.2/7.7 ratio, which is sensitive to the PAH grain-size distribution, decreases with increasing sSFR, suggesting a shift toward relatively larger PAH grains in systems with more intense radiation fields. For the 3.3/11.3 ratio, only two sources are detected, therefore, no clear trend with sSFR can be established. 

Motivated by the increase of 11.3/7.7  with sSFR, we further examine this ratio as a function of the SFR surface density, $\Sigma_{\rm SFR}$. Derived using the effective radius of each galaxy, $\Sigma_{\rm SFR}$ provides a galaxy-averaged measure of the intensity of star formation per unit area and, consequently, a proxy for the mean UV radiation field experienced by the PAH population (Fig.~\ref{fig: SFR surface density vs 11/7}). The positive correlation suggests that the observed PAH processing is not only related to the burstiness of the star formation (high sSFR), but also to how concentrated the star formation is within the galaxy. That is, the 11.3/7.7 PAH ratio is also enhanced in galaxies with higher radiation energy densities, where more UV photons are produced per unit area. In these more intense radiation fields, the small and ionized PAHs that contribute strongly to the 7.7~$\mu$m complex may be preferentially photo-destroyed, while larger and more neutral PAHs, which contribute more efficiently to the 11.3~$\mu$m emission, can survive. In other words, the trend likely reflect differences in the central concentration of star-forming activity within the area defined by the effective-radius of the galaxy: when a larger fraction of star formation is concentrated within smaller areas, PAHs may be exposed to more intense and localized feedback, leading to a more efficient processing of the PAH population. The resulting outcome is the increase of the 11.3/7.7 PAH ratio, mostly due to the preferential slight suppression of the 7.7~$\mu$m feature.

Notably, ASPECS-15, the only AGN-hosting galaxy in the sample, exhibits the highest sSFR, the lowest PAH~6.2/7.7 ($\sim$0.13) and it is one of the source with no 3.3 $\mu m$ detection.  A possible explanation is that the AGN in the PAHSPECS sample is embedded in ISM conditions that are more extreme than those typically found in local AGN hosts, including higher ionization parameter and more intense radiation fields. A larger sample of AGN at cosmic noon is needed to place stronger constraints on the dust properties of this type of sources at high redshift, and to enable a more robust statistical comparison with local galaxies hosting dominant AGN. 

\subsection{The (lack of) 3.3~$\mu$m PAH in cosmic noon galaxies}
\label{3.3 Discussion}
The detected sources, ASPECS-6 and ASPECS-14, show $L_{3.3}/L_{\rm IR}$ values broadly consistent with the nearby-galaxy distribution, while the upper limits for ASPECS-11, ASPECS-15, and ASPECS-C20 do not allow us to establish a systematic suppression of the 3.3~$\mu$m emission. Although the low detection fraction may partly reflect the limited sensitivity of the current observations, it motivates the consideration of physical mechanisms that could weaken this feature in some cosmic-noon galaxies, including enhanced processing of the smallest PAH carriers (see Sect.~\ref{PAH vs sSFR}).
\citet{Donnan26} discusses the possibility that small PAHs are preferentially affected by photo-processing in extreme environments, suggesting that photo-destruction may play a role in shaping the 3.3~$\mu$m emission in the PAHSPECS sample. However, \citet{McKinney25} report detections of the 3.3~$\mu$m PAH feature in 32 of 37 galaxies, with the five non-detections corresponding to sources with the largest AGN contribution to their MIR spectra. This indicates that strong AGN heating may be suppressing the 3.3~$\mu$m PAH feature, either via direct photo-destruction or through dilution due to its contribution to the dust continuum. Indeed, one of the non-detections of the 3.3~$\mu$m feature occurs in ASPECS-15, the only identified AGN in the sample, whose strong radiation field may be destroying the smallest PAH molecules responsible for this feature. Evidence for selective destruction of small PAHs is also well documented in the local Universe. In AGN-dominated galaxies, hard radiation fields have been shown to suppress the smallest PAH grains while larger species remain comparatively resilient (e.g.,\citealp{Inami18, Lai22, Garcia-Bernete22, Garcia-Bernete24, Lofaro26}. Similar effects are, however, not restricted to AGN: intense star-formation-driven radiation fields can likewise modify the PAH population, leading to reduced emission from the smallest grains even in purely star-forming systems (e.g., \citet{Lofaro26}). \citet{McKinney25} reports a reduced relative strength of the 3.3~$\mu$m PAH feature in their sample of (U)LIRGs at cosmic noon, accompanied by 11.3/3.3 ratios that are on average a factor of $\sim$3 higher than those observed in local galaxies. They interpret this trend as evidence for enhanced coagulation in dense ISM environments, where small PAH molecules may grow into larger structures, thereby suppressing the short-wavelength 3.3~$\mu$m emission relative to longer-wavelength neutral features.
However, the PAHSPECS galaxies analyzed in this work lie predominantly on the star-forming main sequence and are not expected to probe similarly extreme density conditions.  
On the other hand, \citet{Lai20} used a sample of nearby star-forming galaxies to show that the 3.3~$\mu$m PAH feature can remain remarkably resilient even in environments characterized by intense radiation fields, suggesting that the smallest PAH grains are not universally destroyed under strong radiation. This highlights that suppression of the 3.3~$\mu$m emission is not an inevitable outcome of elevated radiation intensity alone. Deeper observations of a larger sample of main-sequence galaxies at cosmic noon are required to determine whether the current non-detections are primarily sensitivity limited or whether they reflect genuine variations in the strength of the 3.3~$\mu$m PAH feature. If the latter is confirmed, such variations may point toward differences in ISM conditions relative to nearby star-forming systems, potentially linked to metallicity \citep{Shivaei24,ShivaeiBoo24}, dust composition, or radiation-field geometry.

\subsection{7.7~PAH as an SFR tracer and MIR Diagnostics at Cosmic Noon}

Figure~\ref{fig:SFR_7.7} places the PAHSPECS galaxies on the integrated L$_{7.7}$--SFR plane and shows that they are broadly consistent with the relation defined by nearby dusty galaxies. This agreement is in line with several works in the literature establishing PAH emission and, specifically, the 7.7~$\mu$m complex, as an empirical tracer of star formation across a wide range of environments \citep[e.g.,][]{Brandl06, Smith07, Shipley16, Xie19, Kim24}. At high redshift, \textit{Spitzer}/IRS spectroscopy of luminous $z{\sim}1$--3 systems demonstrated that strong 6--11 $\mu m$~PAH emission is common, and that PAH luminosities broadly track IR-powered star formation, albeit with a large source-to-source diversity associated with ISM conditions and AGN contribution \citep[e.g.,][]{Sajina07, Pope08, Menendez09, Shipley16, Shivaei17}. With \textit{JWST}, direct calibrations and tests of the 7.7~$\mu$m--SFR connection are now emerging from large surveys and resolved nearby templates, supporting the use of 7.7~$\mu$m emission (or its MIRI-band proxies) as first-order SFR indicator while also highlighting second-parameter dependencies such as metallicity and stellar population mix \citep[e.g.,][]{Shivaei17, Ronayne23, Belfiore23, Gregg26}. 

The 7.7 $\mu m$ PAH feature offers a valuable means of locating the cosmic noon galaxies in parameter space and assessing their differences relative to the local IR-luminous population. Fig.~\ref{fig:6.2_EW} shows the L$_{7.7}/\rm L_{\rm IR}$ versus the 6.2~EQW wich decreases when a strong hot-dust continuum dilutes the PAH features and is commonly used to distinguish between star-formation and AGN-dominated mid-infrared spectra \citep[e.g.,][]{Laurent00,Brandl06,Spoon07,Sajina07}. Empirically, thresholds based on 6.2~EQW have been adopted in many \textit{Spitzer} studies to classify sources as AGN-dominated, composite, or star-formation dominated \citep[e.g.,][]{Armus07, Wu10, Sabrina14}.
All PAHSPECS galaxies lie at relatively high equivalent widths, in the range ${\sim}0.5$--$1.4~\mu$m, indicating that their mid-infrared emission is primarily powered by star formation and in agreement with the high equivalent widths observed in local star-forming galaxies \citep{Smith07}.   Their $L_{7.7}/L_{\mathrm{IR}}$ ratios are average with respect to the local galaxy population with a more moderate PAH contribution to $L_{\mathrm{IR}}$ than the most PAH-rich local star-forming galaxies, whose 7.7 $\mu m$ feature can contribute up to 10,\% of their total IR luminosities \citep{Brandl06, Smith07}.
ASPECS-15, the AGN-hosting galaxy, show the lowest $L_{7.7}/L_{\rm IR}$ value and the lowest 6.2~$\mu$m EQW, suggesting an enhanced MIR continuum contribution relative to the PAH emission, possibly linked to the presence of the active nucleus.

\section{Summary and Conclusions}
\label{conclusions}
We have presented JWST/MIRI MRS observations of a sample of five main-sequence star-forming galaxies at $z{\sim}1.1$ drawn from the ASPECS survey in the HUDF. After extracting their integrated spectra from the IFU data cubes, we decomposed the MIR emission using the \texttt{CAFE} fitting tool. We analyzed PAH luminosities and band ratios and investigated their relation to global galaxy properties such as $\rm L_{\rm IR}$, SFR, sSFR, and $\Sigma_{SFR}$. We further compared the integrated PAH properties of the PAHSPECS galaxies to LIRGs from the GOALS survey in order to assess potential systematic differences in ISM conditions between $z{\sim}1$ systems and nearby highly-star-forming galaxies.

The main results of this work can be summarized as follows:
\begin{itemize}

\item Compared to nearby systems, four out of five PAHSPECS galaxies exhibit higher PAH~6.2/7.7 and lower PAH~11.3/7.7, suggesting that the ionized PAH population may be weighted toward smaller grains. The 3.3/11.3 ratio remains poorly constrained because only two sources are detected at 3.3~$\mu$m; therefore, the present data do not allow us to determine whether the neutral PAH population differs systematically in size from that of nearby systems. Nevertheless, our measurements are compatible with the trend toward larger neutral PAHs reported in previous studies of cosmic-noon galaxies. ASPECS-15, the only identified AGN host in the sample, exhibits the lowest PAH~6.2/7.7 and PAH~11.3/7.7 ratios among the PAHSPECS galaxies. These values may indicate both a reduced relative contribution from small PAHs and an enhanced contribution from ionized PAHs, possibly associated with the presence of nuclear activity. Its particularly low PAH~11.3/7.7 ratio places it toward a more highly ionized regime than the local AGN-dominated systems.

\item Within the PAHSPECS sample, 11.3/7.7 shows a positive correlation with sSFR (starburstiness), reflecting selective processing or destruction of small and ionized PAH carriers and the survival of the larger and more neutral ones. This interpretation is supported by the fact that both L$_{6.2}/L_{IR}$ and L$_{7.7}/L_{IR}$ decline with increasing sSFR, i.e. with more intense radiation fields. We also find a positive correlation of the 11.3/7.7 PAH ratio with $\Sigma_{\rm SFR}$, suggesting that PAH processing is linked not only to starburstiness, but also to the surface density of star formation. This trend may also reflect the preferential destruction of small and/or ionized PAHs associated with the 7.7~$\mu$m complex in high-radiation-density environments, since more UV photons are produced per unit area, leaving larger and more neutral PAHs relatively more prominent. 

\item The 3.3~$\mu$m PAH feature is detected in only two out of five galaxies in our sample, while the remaining sources are constrained by upper limits. The limited detection fraction may partly reflect the sensitivity of the current observations; alternatively, it may be associated with enhanced processing of the smallest PAH carriers in harder radiation fields. Deeper observations of larger samples are required to distinguish between these possibilities.

\item The 7.7~$\mu$m PAH luminosity of the PAHSPECS galaxies follows the local $L_{7.7}$--SFR relation, supporting its use as a tracer of star formation at $z\sim1$. The PAHSPECS galaxies span moderate $L_{7.7}/L_{\rm IR}$ values and generally high 6.2~$\mu$m PAH equivalent widths, indicative of MIR emission dominated by star formation. ASPECS-15, the only identified AGN host in the sample, exhibits the lowest EQW$_{6.2}$, possibly reflecting an enhanced MIR continuum contribution associated with nuclear activity.

\end{itemize}

While the present results highlight systematic patterns in the PAH properties of cosmic noon galaxies, the small number of sources limits the statistical significance of these trends. Expanding the sample with additional JWST observations will be essential to find stronger correlations. A natural next step will be to incorporate MIR fine-structure lines into the analysis, allowing us to connect PAH diagnostics with independent probes of ionization and ISM conditions.

\begin{acknowledgements}

This work is based on observations made with the NASA/ESA/CSA James Webb Space Telescope. The data were obtained from the Mikulski Archive for Space Telescopes at the Space Telescope Science Institute, which is operated by the Association of Universities for Research in Astronomy, Inc., under NASA contract NAS 5-03127 for JWST. These observations are associated with program \#5279. Support for program \#5279 was provided by NASA through a grant from the Space Telescope Science Institute, which is operated by the Association of Universities for Research in Astronomy, Inc., under NASA contract NAS 5-03127.

\\ CML acknowledges the research project was supported by the Hellenic Foundation for Research and Innovation (HFRI) under the "2nd Call for HFRI Research Projects to support Faculty Members $\&$ Researchers" (Project Number: 03382).IS acknowledges fundings from the European Research Council (ERC)
DistantDust (Grant No.101117541) and the Atracc\'{i}on de Talento Grant
No.2022-T1/TIC-20472 of the Comunidad de Madrid, Spain. L.A.B. acknowledges support from the Dutch Research Council (NWO) under grant VI.Veni.242.055 (\url{https://doi.org/10.61686/LAJVP77714}).  
K.S. and F.D acknowledge funding support from JWST-GO-5279.002-A. MA is supported by FONDECYT grant number 1252054, and gratefully acknowledges support from ANID Basal Project FB210003,  ANID MILENIO NCN2024_112 and ANID + Vinculaci\'on Internacional + FOVI250261. FG acknowledges support by the French National Research Agency under the contracts WIDENING (ANR-23-ESDIR-0004) and REDEEMING (ANR-24-CE31-2530), as well as by the Actions Thématiques
``Physique et Chimie du Milieu Interstellaire'' (PCMI) of CNRS/INSU, with INC and INP, and ``Cosmologie et Galaxies'' (ATCG) of CNRS/INSU, with INP and IN2P3, both programs being co-funded by CEA and CNES.

\end{acknowledgements}

%

%


\bibliographystyle{aa}

\bibliography{biblio}

\clearpage
\appendix

\section{CAFE fits}
\label{appendix with CAFE fits} 
The following figures show the \texttt{CAFE} spectral energy distribution fits for each ASPECS source in the sample.

\begin{figure*}[t]
    \centering
    \includegraphics[width=0.9\textwidth]{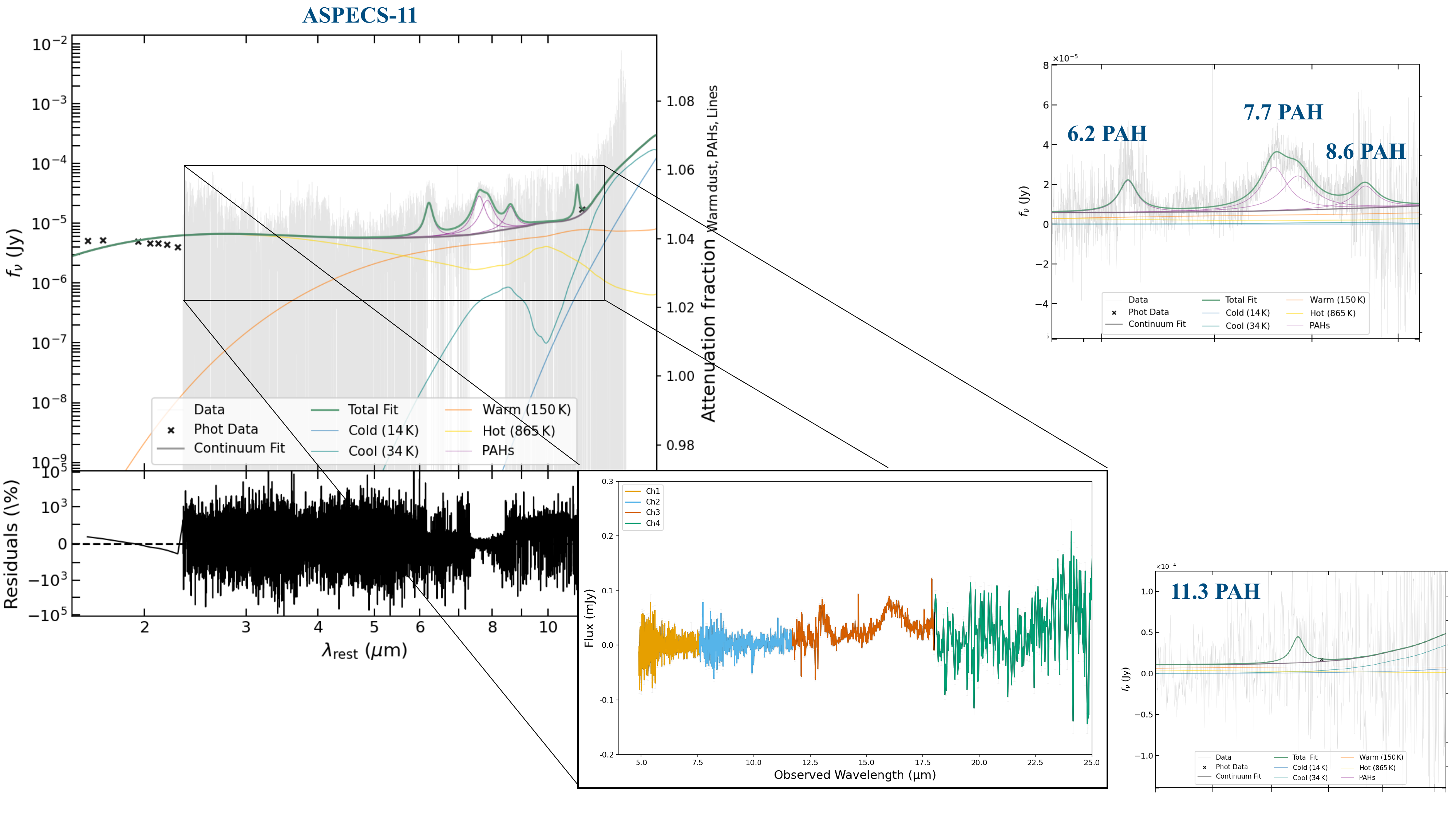}
    \caption{ASPECS-11: Spectral fit using the full JWST/MIRI MRS wavelength coverage from 5 to 28~$\mu$m, together with the available photometric data. The symbols follow Fig.~\ref{fig: CAFE fit aspecs6}.}
    \label{fig: CAFE fit aspecs11}
\end{figure*}

\begin{figure*}[]
	\centering
		\includegraphics[width=0.9\textwidth]{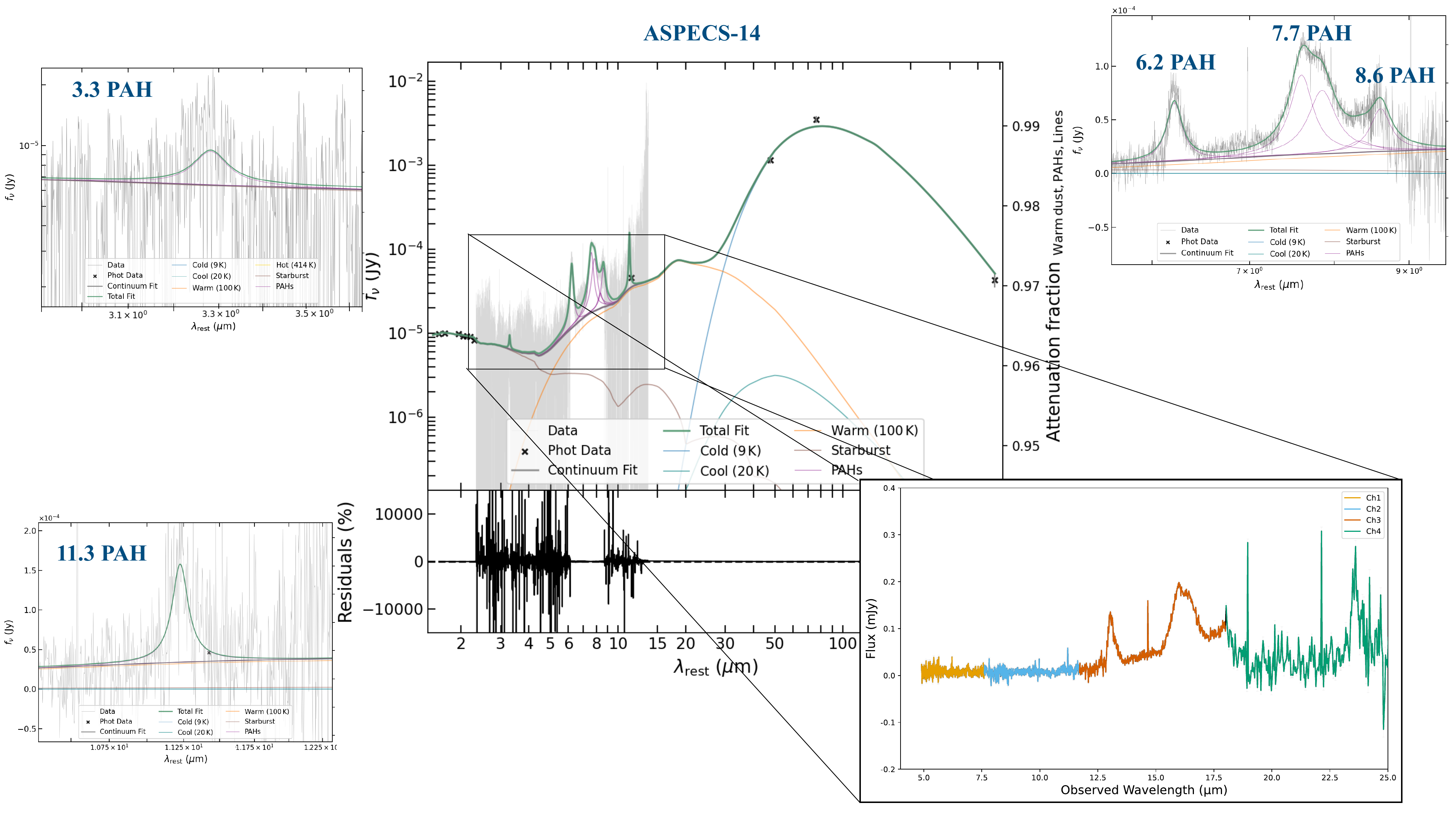}
		\caption{ASPECS-11: Spectral fit using the full JWST/MIRI MRS wavelength coverage from 5 to 28~$\mu$m, together with the available photometric data. The symbols follow Fig.~\ref{fig: CAFE fit aspecs6}}
        \label{fig: CAFE fit}
\end{figure*}

\begin{figure*}[]
	\centering
		\includegraphics[width=1\textwidth]{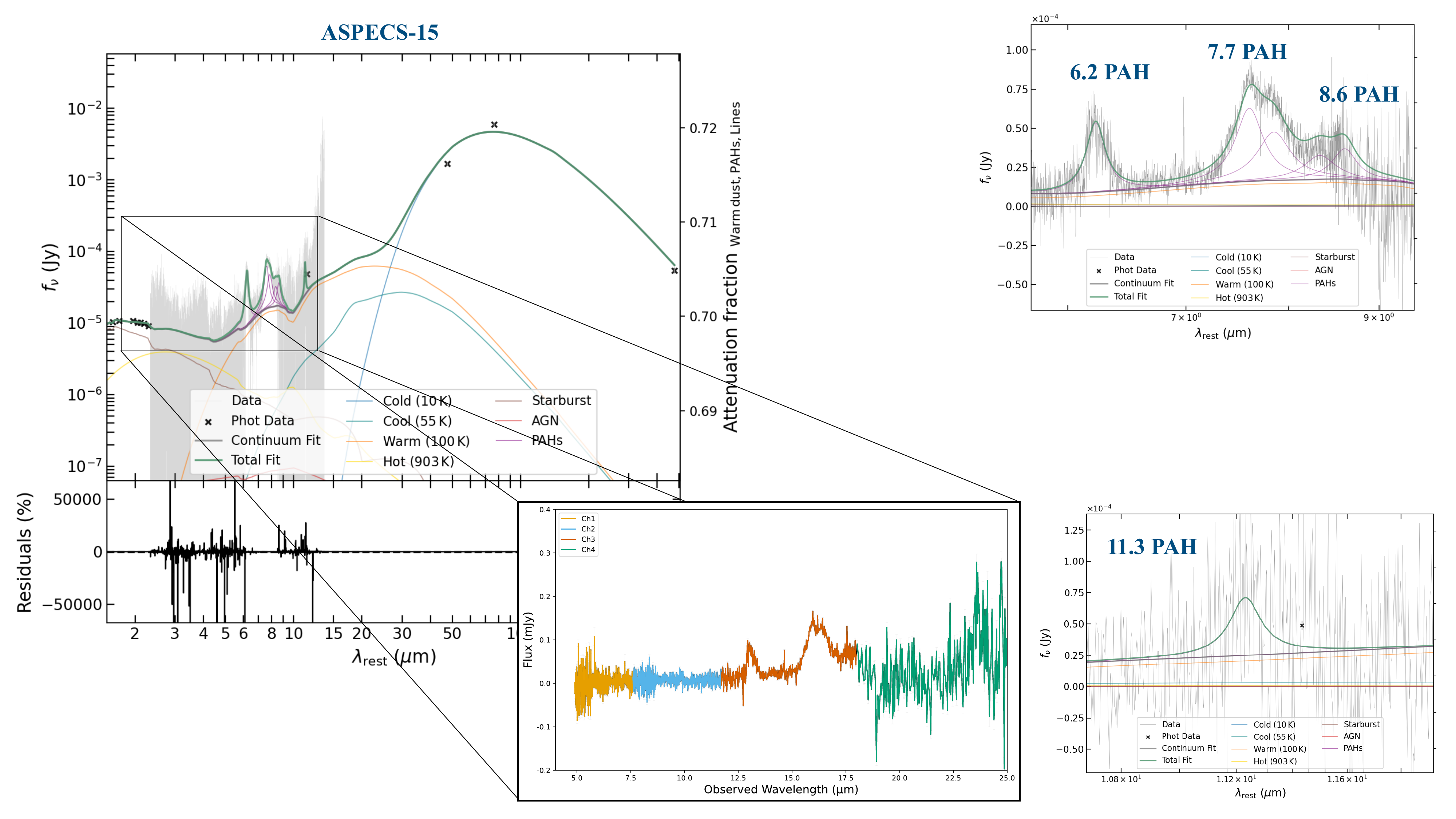}
		\caption{ASPECS-15: Spectral fit using the full JWST/MIRI MRS wavelength coverage from 5 to 28~$\mu$m, together with the available photometric data. The symbols follow Fig.~\ref{fig: CAFE fit aspecs6}}
        \label{fig: CAFE fit}
\end{figure*}

\begin{figure*}[]
	\centering
		\includegraphics[width=1\textwidth]{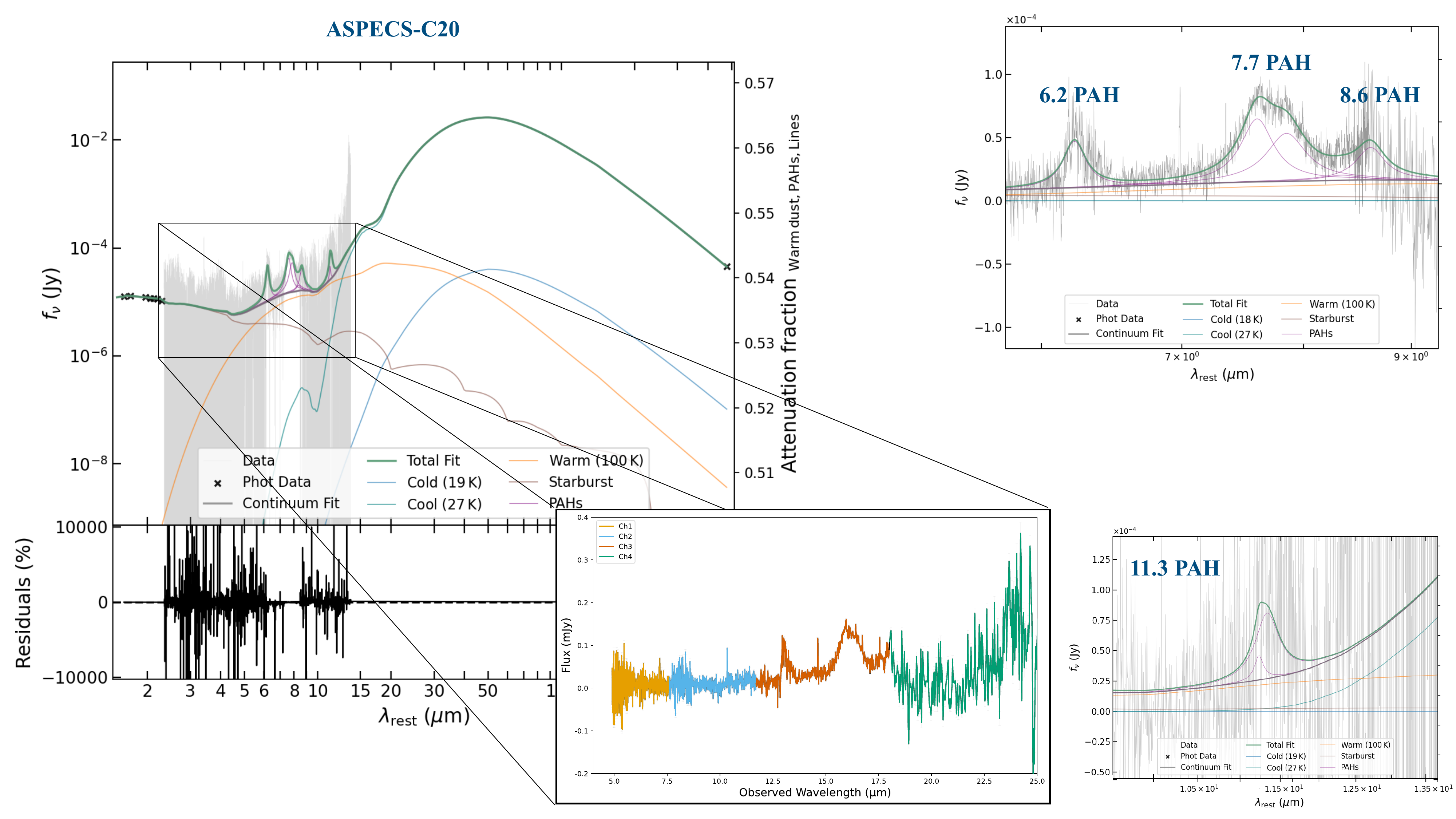}
		\caption{ASPECS-20: Spectral fit using the full JWST/MIRI MRS wavelength coverage from 5 to 28~$\mu$m, together with the available photometric data. The symbols follow Fig.~\ref{fig: CAFE fit aspecs6}}
        \label{fig: CAFE fit}
\end{figure*}

\end{document}